\documentclass[aps,prl,reprint,superscriptaddress]{revtex4-2}

\usepackage{textcomp}
\usepackage{graphicx}
\usepackage{amsmath}
\usepackage{amssymb}
\usepackage{siunitx}
\usepackage[colorlinks=true, citecolor={black}, urlcolor={black}, linkcolor = {black}]{hyperref}
\makeatletter
\usepackage{cleveref}
\Crefname{figure}{Fig.}{Figs.}
\crefname{figure}{Fig.}{Figs.}

\newcommand{\fmarki}{*}
\newcommand{\fmarkii}{\ensuremath{\dagger}}
\newcommand{\fmarkiii}{\ensuremath{\ddagger}}
\newcommand{\fmarkiv}{\ensuremath{\mathsection}}
\newcommand{\fmarkv}{\ensuremath{\mathparagraph}}
\newcommand{\fmarkvi}{\ensuremath{\|}}
\newcommand{\fmarkvii}{**}
\newcommand{\fmarkviii}{\ensuremath{\dagger\dagger}}
\newcommand{\fmarkix}{\ensuremath{\ddagger\ddagger}}
\def\@fnsymbol#1{{\ifcase#1\or \fmarki\or \fmarkii\or \fmarkiii\or \fmarkiv\or \fmarkv\or \fmarkvi\or \fmarkvii\or \fmarkviii\or \fmarkix \else\@ctrerr\fi}}
\makeatother

\renewcommand{\fmarki}{\ensuremath{\dagger}}
\renewcommand{\fmarkii}{*}

\def\uppi~{$\mathrm{\pi}$}

\newcommand{\fl}[1]{\textbf{(#1)}}

\newcommand{\var}[2]{$#1_{\mathrm{#2}}$}

\newcommand{\ABS}[1]{$V_{\mathrm{ABS}}^{\mathrm{(#1)}}$}

\newcommand{\ket}[1]{$\vert #1\rangle$}

\newcommand{\ua}{\uparrow}
\newcommand{\da}{\downarrow}

\begin{document}

\title{Probing ground-state degeneracies of a strongly interacting Fermi-Hubbard model
with superconducting correlations.}

\author{Sebastiaan L. D. ten Haaf}
\altaffiliation{These authors contributed equally to this work.}
\affiliation{QuTech and Kavli Institute of Nanoscience, Delft University of Technology, Delft, 2600 GA, The Netherlands}

\author{Sebastian Miles}
\altaffiliation{These authors contributed equally to this work.}
\affiliation{QuTech and Kavli Institute of Nanoscience, Delft University of Technology, Delft, 2600 GA, The Netherlands}

\author{Qingzhen Wang}
\affiliation{QuTech and Kavli Institute of Nanoscience, Delft University of Technology, Delft, 2600 GA, The Netherlands}

\author{A. Mert Bozkurt}
\affiliation{QuTech and Kavli Institute of Nanoscience, Delft University of Technology, Delft, 2600 GA, The Netherlands}

\author{Ivan Kulesh}
\affiliation{QuTech and Kavli Institute of Nanoscience, Delft University of Technology, Delft, 2600 GA, The Netherlands}

\author{Yining Zhang}
\affiliation{QuTech and Kavli Institute of Nanoscience, Delft University of Technology, Delft, 2600 GA, The Netherlands}

\author{Christian G. Prosko}
\affiliation{QuTech and Kavli Institute of Nanoscience, Delft University of Technology, Delft, 2600 GA, The Netherlands}

\author{Michael Wimmer}
\affiliation{QuTech and Kavli Institute of Nanoscience, Delft University of Technology, Delft, 2600 GA, The Netherlands}

\author{Srijit Goswami}\email{s.goswami@tudelft.nl}
\affiliation{QuTech and Kavli Institute of Nanoscience, Delft University of Technology, Delft, 2600 GA, The Netherlands}

\begin{abstract}
The Fermi-Hubbard model and its rich phase diagram naturally emerges as a description for a wide range of electronic systems.
Recent advances in semiconductor-superconductor hybrid quantum dot arrays have allowed to realize degenerate quantum systems in a controllable way, e.g., allowing to observe robust zero-bias peaks in Kitaev chains, indicative for Majorana bound states.
In this work, we connect these two domains.
Noting the strong on-site Coulomb repulsion within quantum dots, we study small arrays of spinful hybrid quantum dots implemented in a two-dimensional electron gas.
This system constitutes a Fermi-Hubbard model with inter-site superconducting correlations.
For two electronic sites, we find robust zero-bias peaks indicative of a strongly degenerate spectrum hosting emergent Majorana Kramers pairs or $\mathbb{Z}_3$-parafermions.
Extending to three sites, we find that these spinful systems scale very differently compared to spinless Kitaev chains. 
When the sweet-spot conditions are satisfied pairwise, we find that the ground state degeneracy of the full three-site system is lifted. 
This degeneracy can be restored by tuning the superconducting phase difference between the hybrid segments. 
However, these states are not robust to quantum dot detuning.
Our observations are a first step towards studying degeneracies in strongly interacting Fermi-Hubbard systems with superconducting correlations.
\end{abstract}

\maketitle

\section{Introduction}
The Fermi-Hubbard model is a fundamental model for strongly correlated electron systems, capturing the competition between electron kinetic energy and on-site Coulomb repulsion~\cite{hubbard_model, hubbard_model_review}.
Despite its seeming simplicity, the model exhibits a rich phase diagram, including metallic, insulating, magnetic, and superconducting phases~\cite{hubbard_model_review, lee_doping_2006}.
This fundamental importance has motivated many experimental efforts to realize the Fermi-Hubbard model in controllable platforms such as ultra-cold atoms in optical lattices~\cite{esslinger_fermi-hubbard_2010}, trapped ions~\cite{johanning_quantum_2009}, and superconducting circuits~\cite{schmidt_circuit_2013, houck_-chip_2012}.
Semiconductor quantum dots are another natural platform, due to the intrinsic strong electron-electron interactions in the form of on-site Coulomb repulsion~\cite{yang_generic_2011}.
Recent experiments have studied different aspects of the Fermi-Hubbard model using small quantum dot arrays~\cite{hensgens_quantum_2017, byrnes_quantum_2008}.\newline
Semiconductor quantum dots were also put forth as a platform to implement a non-interacting model exhibiting topological superconductivity - the Kitaev chain~\cite{Kitaev2001, Sau2012, Leijnse2012, Fulga2013}.
Recently, short Kitaev chain were successfully implemented using small arrays of quantum dots coupled by superconducting hybrid segments~\cite{Dvir2023, tenHaaf2024, bordin2024, haaf2025}.
In addition to normal electron hopping between quantum dots, the hybrid superconducting segments mediate Cooper pair splitting processes, effectively implementing a p-wave pairing term between neighbouring quantum dots~\cite{CXLiu2022, Tsintzis20222}.
Though the charging energy is still the largest energy scale in these devices~\cite{tenHaaf2024}, the effects of interactions are minimized by a finite Zeeman splitting on the dots, effectively suppressing interactions between different spin species.\newline\newline
In this work, we combine these two research directions, and study a superconducting extension of the Fermi-Hubbard model using quantum dot arrays.
To this end, we experimentally implement short chains of quantum dots coupled by superconducting hybrid segments, as used to implement Kitaev chains~\cite{Dvir2023, tenHaaf2024, Zatelli2024}, but now in the absence of a magnetic field.
In this regime, the effect of electron interactions on the quantum dots is restored.
This effectively yields a one-dimensional Fermi-Hubbard model with an additional nearest-neighbour, spatially-odd pairing term~\cite{bozkurt2025}.
This model was recently shown theoretically to feature peculiar ground state degeneracies due to the interplay between strong interactions and superconducting pairing.
In particular, a two-site system was predicted to exhibit three-fold degenerate spectrum due to correlations \cite{bozkurt2025}, linked to the emergence of strong zero modes~\cite{Fendley2012, Alicea2016} protected by a $\mathbb{Z}_3$ parity operator.

We implement devices with effectively two and three interacting sites on InSbAs 2DEGs with epitaxial Al and study their low energy spectra through conductance measurements.
Our experiments reproduce the main results of the theoretical predictions in Ref.~\cite{bozkurt2025}: a stable ground state degeneracy is observed in the two-site chains, which is lifted when extending to three sites. 
Additionally, we find in both theory and experiment that introducing a superconducting phase difference between the hybrid QDs allows to tune also a three-site system to a strongly degenerate configuration.
The striking agreement between experiment and theory suggests that the quantum dot platform is well suited to explore the interplay between strong correlations and superconductivity in extended Fermi-Hubbard type models.

\subsection*{The spinful QD chain model}
A chain of QDs coupled via superconductors can be described by a Fermi-Hubbard Hamiltonian with an additional nearest-neighbour superconducting term:

\begin{align} \label{eq:3_site_effective}
    H &= H_{\mathrm{D}} + H_{\mathrm{U}} + H_{\mathrm{ECT}} + H_{\mathrm{CAR}}.
\end{align}

Here, $H_{\mathrm{D}} = \sum_{i} \mu_{i} (n_{i,\ua}+n_{i,\da})$ models the QDs chemical potential $\mu_i$, while $ H_{\mathrm{U}} = \sum_{i} U_i n_{i,\ua}n_{i,\da}$ models the on-site Coulomb repulsion.
The superconducting segment hosts gate-tunable Andreev bound states, that generate two virtual transport processes coupling the neighbouring QDs~\cite{CXLiu2022, WangG2022,Bordin2024threesiteCAR,Bordin2023,WangQ2023}.
Elastic co-tunneling (ECT) facilitates hopping of an electron between two sites. 
We model this process as $H_{\mathrm{ECT}}=\sum_{i} t^{(i,i+1)} (c_{i,\uparrow}^\dagger c_{i+1,\uparrow} + c_{i,\downarrow}^\dagger c_{i+1,\downarrow}) + h.c.$ between sites $i$ and $j$.
Crossed Andreev reflection (CAR), modelled by $H_{\mathrm{CAR}}=\sum_{i} \Delta^{(i,i+1)} (c_{i,\uparrow}^\dagger c_{i+1,\downarrow}^\dagger - c_{i,\downarrow}^\dagger c_{i+1,\uparrow}^\dagger) + h.c. $, facilitates a pairing interaction via the creation or breaking of Cooper pairs in a superconductor.
Notably, due to only including nearest-neighbour couplings, spin-orbit interactions can be gauged away (see Methods) and hence we only include spin-preserving processes.\newline
In the devices of interest the typical interdot coupling amplitudes are on the order of \SI{20}{\micro\volt}, while typical charging energies are on the order of 1-\SI{2}{\milli\volt}, such that Coulomb repulsion is the largest energy scale in the system (\cref{fig:S1_charact_devA}).
We therefore restrict ourselves to the $U=\infty$ limit to compare experiments and theory in the main text. 
To capture this, we project out the double occupancy restricting us to an effective basis \{\ket{0}, \ket{\mathord{\ua}}, \ket{\mathord{\da}}\} on each QD.
A summary of more extensive modelling that includes the nearest neighbour processes can be found in Methods.
For two QD sites, Refs.~\cite{OBrien2015, Wright2013} have shown that the system features a `sweet spot' at $\mu_1=\mu_2 =0$ and $t^{(1,2)}=\sqrt{2}\Delta^{(1,2)}$.
Here, the ground state of the system is threefold degenerate and hosts Majorana Kramers pairs~\cite{bozkurt2025}, which are stable against detuning of individual QDs.
Below, we start by characterising such `sweet spots' in a two QD system experimentally and study the evolution of the low-energy spectrum when adding a third QD. 

\begin{figure}
\centering
    \includegraphics[width = 0.5\textwidth]{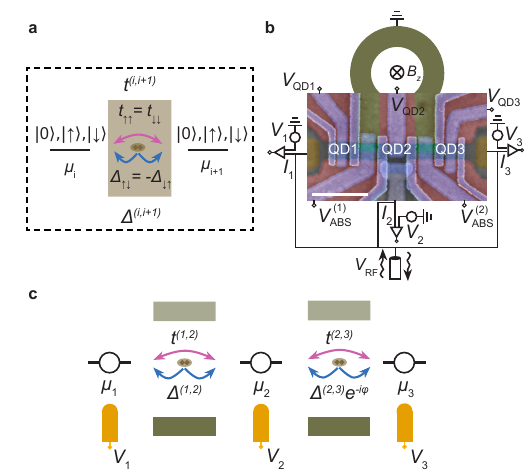}
    \caption{\textbf{Model and device schematics} \fl{a} Schematic of the nearest-neighbour interactions for two spinful fermionic sites coupled via a superconductor. We assume equal tunneling rates for spin-up and spin-down such that $t^{(i,j)}=t^{(i,j)}_{\downarrow\downarrow}=t^{(i,j)}_{\uparrow\uparrow}$ and $\Delta^{(i,j)}=\Delta^{(i,j)}_{\ua\da}=-\Delta^{(i,j)}_{\da\ua}$, where the sign is due to time reversal symmetry.
    To implement the $U=\infty$ limit, the double occupied state is disallowed. 
    \fl{b} False colour scanning electron micrograph of a copy of device A. 
    Two aluminium strips are connected via a continuous loop with a radius of~\SI{10}{\micro\meter} (not drawn to scale). 
    Scale bar is \SI{500}{\nano\meter}. 
    \fl{c} Schematic representation of the three site device, highlighting the parameters relevant for the theoretical simulations.
    }
    \label{fig:Fig1}
\end{figure}

\subsection*{Device and measurement set-up}
The device used to obtain the results in the main text is shown in \cref{fig:Fig1}b, containing three quantum dots and two superconducting regions.
The two SC strips (green) are connected in a continuous loop and kept grounded. 
An external magnetic field \var{B}{z} controls the superconducting phase difference between the strips.
Andreev bound states (ABSs) are induced in the regions proximitized by the SCs (\cref{fig:S1_charact_devA}).
The energies of the ABSs in the left and right hybrid sections are controlled by the voltages denoted \ABS{1} and \ABS{2}.
Three large gates (red) are used to confine a quasi-1D channel. 
Narrow finger gates (blue), separated by a dielectric layer, are used to confine three QDs and control their electrochemical potentials (\var{V}{QDi}).
Each QD can be probed by an Ohmic lead, which enable recording the device conductance in two separate ways.
Applied voltages (\var{V}{1}, \var{V}{2} and \var{V}{3}) and measured currents (\var{I}{1}, \var{I}{2} and \var{I}{3}) are used to measure the local differential conductance in each lead ($G_{ii}=\frac{dI_{i}}{dV_{i}}$).
In addition, each lead is connected to an off-chip resonator, which allows for measuring the RF-reflectometry response, denoted \var{\tilde{S}}{i} (for details, see Methods and \cref{fig:S2_RF_processing}).
This signal is linearly proportional to the device conductance in the regime of interest~\cite{reilly2007fast, jung2012radio,razmadze2019radio}, explicitly shown in Ref.~\cite{kulesh2025} for this device.

\begin{figure}
\centering
    \includegraphics[width = 0.5\textwidth]{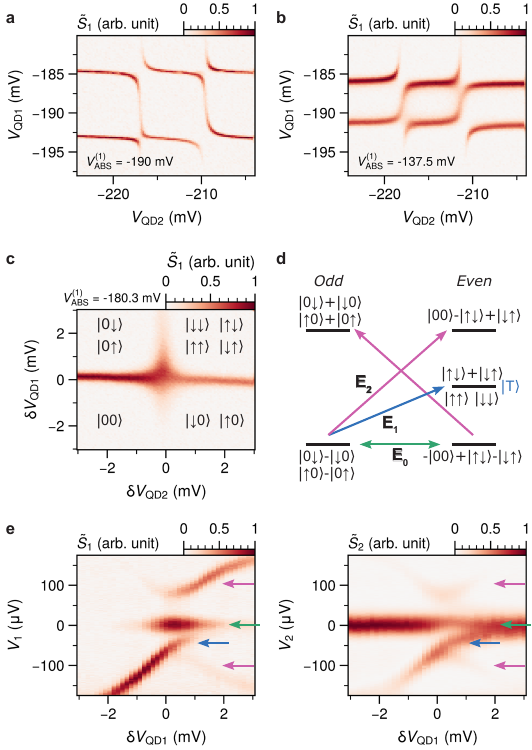}
    \caption{\textbf{The two-site sweet spot.}
    Tuning the coupling between QD1 and QD2, with QD3 tuned to a Coulomb blockaded regime.
    \fl{a} Exemplar CSD in an ECT dominated regime (\ABS{1}~=~\SI{-190}{\milli\volt}), showing four anti-diagonal avoided crossings.
    \fl{b} Exemplar CSD in a CAR dominated regime (\ABS{1}~=~\SI{-137.5}{\milli\volt}), showing four diagonal avoided crossings.
    Examples were chosen to showcase the change in connectivity can arise in all 4 quadrants of the CSDs. 
    More general CSDs are shown in \cref{fig:S3_extended_CSDs_devA}.
    \fl{c} Close up CSD of the bottom left quadrant, at an intermediate value of \ABS{1}~=~\SI{-180.3}{\milli\volt}.
    \fl{d} Energy level diagram for the two parity sectors at the sweet spot (detailed in the main text). 
    Coherence factors have been omitted for brevity. 
    \fl{e} Tunnelling spectroscopy measured when detuning \var{V}{QD1}, with \var{V}{QD2} set on resonance. 
    Coloured arrows correspond to transitions indicated in (d).
    Reproduction of results in a separate device is shown in \cref{fig:S5_devBcharacterisation}-\ref{fig:S7_reproduce_spectra_devB}.
    }
    \label{fig:Fig2}
\end{figure}

\section*{Results}
\subsection{The two-site chain}

After forming three QDs through electrostatic tuning of the finger gates, we focus first on the coupling between QD1 and QD2, keeping QD3 away from Coulomb resonance.
The key to generating stable zero-energy modes in this two-QD system, widely studied at finite magnetic fields~\cite{Leijnse2012,Tsintzis20222,Dvir2023,Zatelli2024, tenHaaf2024}, is that the simultaneous presence of ECT and CAR creates a competition between different parity ground-states.
ECT couples states with the same total particle number, whereas CAR couples states whose particle number differs by two. 
Notably, this competition can be present also without an external magnetic field~\cite{Scherubl2019}.
To observe this, we study charge stability diagrams (CSDs) across a single orbital in each QD through lead reflectometry.
When ECT dominates, diagonal avoided crossings are expected, as shown in an exemplary measurement in \cref{fig:Fig2}a.
Control over \ABS{1} affects the underlying CAR and ECT processes, which can drastically change the CSD.
Changing \ABS{1} by $\approx$~\SI{50}{\milli\volt}, the connectivity in the entire CSD changes, showing now only antidiagonal avoided crossings (\cref{fig:Fig2}b), indicating CAR has become the dominant process.
This continuous control over the connectivity ensures that, for each pair of resonances, a `sweet-spot' exists where the avoided crossing disappears.
There, zero-energy modes are expected to arise that are resilient against detuning individual QDs (local perturbations) and protected to at least second-order against joint detuning of all QDs (global perturbations, cf.~\cite{bozkurt2025}).
Zooming in on the bottom-left pair of transitions and fine-tuning \ABS{1}, we find such a `sweet spot' in \cref{fig:Fig2}c.
\newline

To understand the density of states, we consider the energy level diagram of the effective model at the sweet spot ($t^{(1,2)}=\sqrt{2}\Delta^{(1,2)}$, \var{\mu}{i}~=~0) in \cref{fig:Fig2}d, following Ref.~\cite{bozkurt2025}.
The spectrum is strongly degenerate, i.e. all energy manifolds feature a three-fold degeneracy.
This structure results in single-electron transitions from the ground-states at three possible energies, labelled by the coloured arrows.
We measure these transitions experimentally through lead-reflectometry of lead 1 and lead 2 and sweep \var{V}{QD1}.
The three expected transitions can be clearly identified~(\cref{fig:Fig2}e).
Most trivially, we observe excitations symmetrically around zero energy which disperse with \var{V}{QD1}, denoted as $E_2$ (pink).
Secondly, we see stable zero-bias conductance peaks (green) reflecting the zero-energy excitations (\var{E}{0}) between the degenerate even and odd ground-states.
These excitations can be understood as arising from Kramers pairs of Majorana modes, as detailed in~\cite{bozkurt2025}.
It is important to note that they are not expected to be localized on either of the QDs, but are nevertheless robust against perturbing the chemical potential of a single QD (see also \cref{fig:S4_right_PMM}). 
Lastly, we find a transition appearing only at negative bias (blue). 
This transition, denoted \var{E}{1}, is a signature of the presence of the triplet states \ket{\mathrm{T}} in the even parity subspace~\cite{Scherubl2019}, breaking particle hole symmetry in the conductance spectrum.
Notably, this appears to be the only feature that distinguishes this system from the two-site system at finite magnetic field~\cite{Dvir2023,Zatelli2024,tenHaaf2024}.
The particle or hole-like nature of the feature depends on the charge configuration of the QDs, which we demonstrate in \cref{fig:S7_reproduce_spectra_devB}.
The presence of these triplet states is expected to be a detrimental factor when scaling the system beyond two sites~\cite{bozkurt2025}, which we study below.

\begin{figure}
\centering
    \includegraphics[width = 0.5\textwidth]{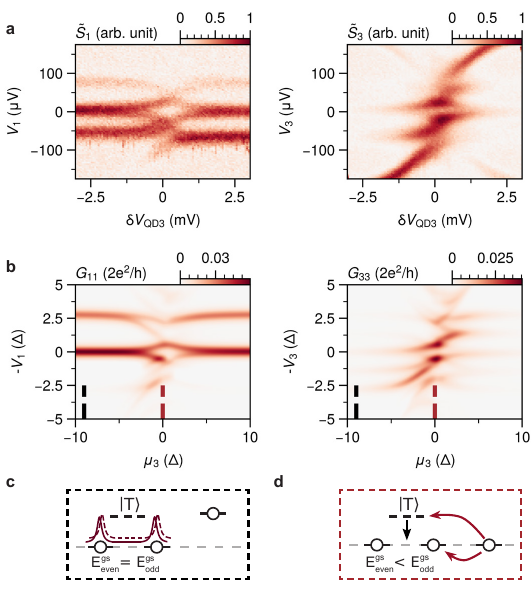}
    \caption{\textbf{Scaling from two to three sites.} Measurements obtained with both pairs of QDs individually tuned to a sweet spot configuration. \fl{a} Spectroscopy of the left (\var{S}{1}) and right (\var{S}{3}) while sweeping \var{V}{QD3}. Once QD3 is on resonance, the ZBP on both sides is observed to split.
    \fl{b} Numerically calculated conductance with (settings), showing similar behaviour.
    To understand this, we consider schematically the situation when QD3 is \fl{c} off and \fl{d} on resonance.
    When QD3 is off resonance, the triplet subspace is uncoupled from the rest of the system and a pair of spatially overlapping Majorana Kramers pairs is present on both QDs. 
    Adding the third QD couples to both zero energy modes and allows the triplet states to participate, favouring the even ground-state in energy (see Ref.~\cite{bozkurt2025} for further details).
    }
    \label{fig:Fig3}
\end{figure}

\subsection{Scaling from two to three sites}
At finite magnetic fields, the conditions that give rise to stable zero modes in the two-site chain can be extended to obtain stable zero modes in longer chains~\cite{Sau2012}. 
By satisfying the `sweet spot' condition between all pairs of neighbouring QDs and setting all QDs on resonance, stable ZBPs were expected on the outermost sites~\cite{Dourado2025, cxliu2025}, which was recently demonstrated experimentally in three-site systems~\cite{bordin2024,haaf2025}.
As shown in~\cite{jermyn2014}, this is not necessarily true for strong zero modes however: the additional states added to the Hilbert space can introduce couplings which split the required strong degeneracy of the levels.
In Ref.~\cite{bozkurt2025}, a crucial part for the existence of a strong degeneracy is that the triplet states in the excited manifold decouple from the rest of the spectrum.
This breaks down when turning on the coupling to the third site despite tuning $t,\Delta$ to the same values.
To demonstrate this, we first repeat the procedure demonstrated in \cref{fig:Fig2} for QD2 and QD3, with QD1 off resonance, to obtain a $t^{(2,3)}=\sqrt{2}\Delta^{(2,3)}$ sweet spot for the right QD pair (\cref{fig:S4_right_PMM}).
With either of the outer QDs off-resonance, the remaining two QDs thus host a pair of ZBPs.
Next, we bring all three QDs on resonance and perform finite-bias spectroscopy measurements.
\cref{fig:Fig3}a shows RF-spectroscopy measurements of \var{\tilde{S}}{1} and \var{\tilde{S}}{3}, when detuning \var{V}{QD3}.
We observe that the ZBPs split in energy as soon as \var{\delta V}{QD3} approaches resonance.
This is in stark contrast to the two-site systems in isolation, which are resilient against local perturbations (\cref{fig:Fig2}), as well as to the behaviour observed for three-site systems at finite magnetic fields~\cite{haaf2025,bordin2024}. 
We find the behaviour closely matches the numerical simulations in \cref{fig:Fig3}c.
The characteristic diamond shape of the splitting is reminiscent of the behaviour expected when coupling a quantum dot to an overlapping pair of Majoranas~\cite{prada2017,Clarke_2017,bordin2025pradaclarke}.
In that context, this measurements highlights the expected non-locality of the Majorana Kramers pairs in this system: even at the sweet-spot for two-sites, the zero-energy modes are not localized on either QD.
We illustrate this interpretation in Figs.~3c,d: what were two delocalized but isolated Majorana Kramers pairs in the two-site system (c) ) begin hybridizing through processes involving the previously uncoupled triplet states (d) ).

\begin{figure}[t!]
\centering
    \includegraphics[width = 0.5\textwidth]{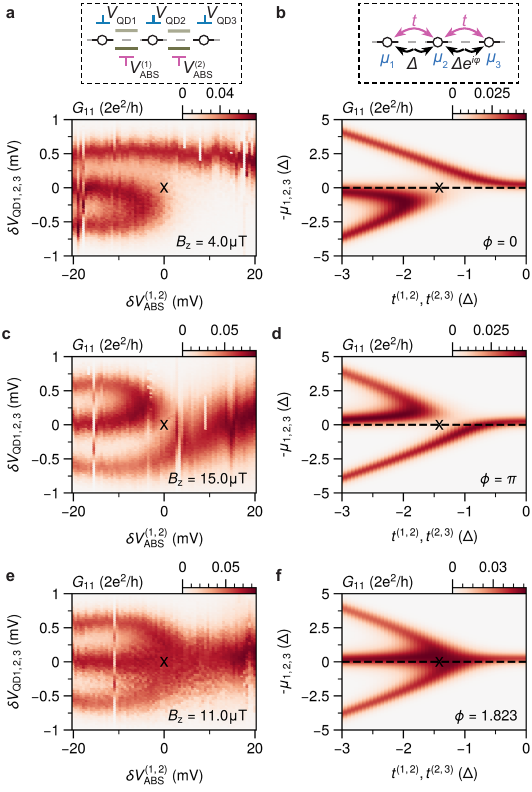}
    \caption{\textbf{Superconducting phase as tuning knob.} To understand where zero-energy modes arise in the full three-site system, a larger parameter space is explored (procedure detailed in \cref{fig:S8_full_data_figure_4}). 
    For each QD pair, sweet spot values are obtained, denoted as $\delta$\ABS{i}~=~0.
    Zero-bias conductance measurements are performed upon simultaneously varying \var{\delta V}{QDi} versus sweeping \ABS{i} around their sweet-spots.
    \fl{a} Measurement obtained for $B_z$~=~\SI{4}{\micro\tesla}, corresponding roughly to $\phi$~=~0. \fl{b} We compare this to a numerical simulation, where all $\mu$ and all $t$ parameters are swept together, for fixed $\Delta$. 
    \fl{c-f} We repeat the measurements and simulations for $B_z$~=~\SI{15}{\micro\tesla} (c), corresponding rouhgly to $\phi$~=~$\pi$ (d) and  (e) $B_z$~=~\SI{11}{\micro\tesla} corresponding to (f) $\phi$~$\approx$~1.82.
    }
    \label{fig:Fig4}
\end{figure}

\subsection{Superconducting phase control}
We turn our attention to the final tunable parameter: the phase difference between the superconductors.
In a two-site system, this phase difference can be gauged away and therefore does not affect the energy spectra. 
This is no longer the case when a third site is added~\cite{Sau2012, Fendley2012, denNijs1983, Ostlund1981}.
Instead, the phase breaks time reversal and inversion symmetry.
Sweeping the phase, it has been predicted to be possible to recover strong zero modes~\cite{Fendley2012}, albeit with reduced stability~\cite{jermyn2014}.
We look for ground-state degeneracies by measuring zero-bias conductance in the parameter space around the \ABS{1} and \ABS{2} values that corresponded to two-site sweet-spots.
Sweeping all three QDs around their charge degeneracy points simultaneously and varying both \ABS{1} and \ABS{2}, we map out the conditions that give rise to zero-bias conductance when $\phi$~=~0, shown in \cref{fig:Fig4}a.
The `x' marks the point in parameter space where ZBPs arise in the isolated two-site systems (as in \cref{fig:Fig2}).
In accordance with \cref{fig:Fig3}, there is now no zero-bias conductance at this point in the three-site system.
We compare the measurement to calculated conductance from the effective model, when all \var{\mu}{i} are varied simultaneously versus $t^{(1,2)}=t^{(2,3)}$, for fixed $\Delta^{(1,2)}=\Delta^{(2,3)}$ (\cref{fig:Fig4}b).
Next, we apply a small out-of-plane field \var{B}{z} through the superconducting loop corresponding to a $\pi$-phase and repeat the measurement, shown in \cref{fig:Fig4}c. 
Interestingly, this results in a mirrored pattern that again matches well with the theory (\cref{fig:Fig4}d).
While at $\pi$-phase we still do not yield a degeneracy at the two-site sweet spot parameters (marked by `x'), the evolution from \cref{fig:Fig4}a implies this should be possible to achieve when fine-tuning the SC phase.
By sweeping the phase for the degeneracy to cross $\mu_i=0$ at $\delta V^{(1,2)}_{\mathrm{ABS}}=0$, we arrive at the measurement shown in \cref{fig:Fig4}e.
Theoretically, we find this corresponds to the exact phase of $\phi=\arccos(-1/4)$, shown in \cref{fig:Fig4}f (see Methods).
At this angle, the system in fact exhibits a three-fold (strongly) degenerate spectrum at the point $\mu=0$, $t=\sqrt{2}\Delta$.
It should be noted that temperature broadening in the experiment results in zero-bias conductance in a larger region, while theory predicts a single point in parameter space.
Going further, we find this point lies along a line in the $\phi-t$ parameter space, which we address in more detail below.

\subsection{Strong degeneracies through control over the superconducting phase} \label{sec:magic_line_discussion}
In \cref{fig:Fig4} we have observed that tuning the superconducting phase difference yields configurations for which the degeneracy lines cross $\mu_i$~=~0.
When $t^{(i,j)}=\sqrt{2}\Delta^{(i,j)}=$ (the two-site sweet-spot), the phase at which this crossing occurs is $\phi=\arccos(-1/4)$, where the entire spectrum is found to be strongly degenerate.
We find that by allowing $\Delta^{(i,j)}\neq t^{(i,j)}$, there exists a whole line in parameter space for which the spectrum is strongly, threefold degenerate.
Fixing $\mu_i=0$ and calculating the spectrum analytically (see Methods), we obtain the condition for degeneracy to be:
\begin{align} \label{eq:magic_line}
    t^{(i,j)}=\sqrt{-\frac{1}{2\cos(\varphi)}}\Delta^{(i,j)},
\end{align}
\begin{figure}[t!]
\centering
    \includegraphics[width = 0.5\textwidth]{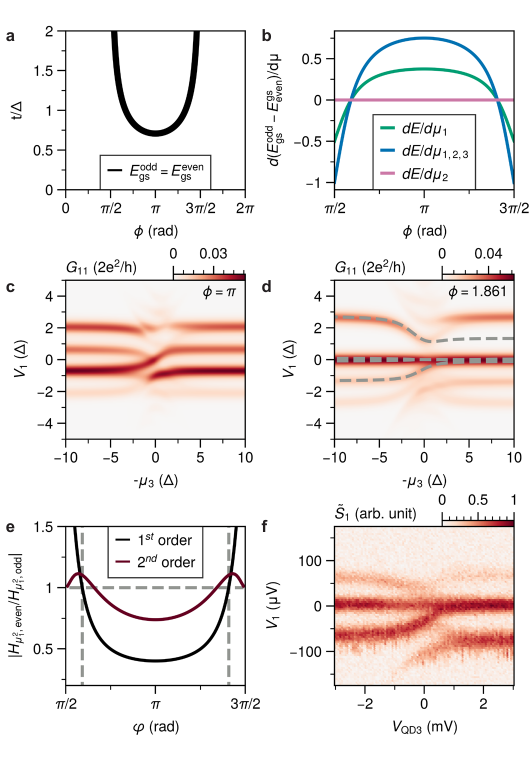}
    \caption{\textbf{Strongly degenerate ground states in the three-site chain.} \fl{a} Analytical solution yielding $E_{\mathrm{gs}}^{\mathrm{odd}}=E_{\mathrm{gs}}^{\mathrm{even}}$ for the three-site chain when $\mu_i=0$ (given by Eq.~\ref{eq:magic_line}), resulting in a strongly degenerate spectrum. 
    \fl{b} Derivative of the energy difference between the odd and even ground states with respect to detuning $\mu$. 
    The derivatives vanish at two specific points. 
    The significance can be seen when comparing the conductance spectrum of detuning \var{\mu}{3} at \fl{c} $\phi$~=~$\pi$ and \fl{d} $\phi$~=~1.861.
    The numerical energy spectrum is overlaid in (d) to highlight that a small energy splitting still arises at larger detuning of $\mu_3$.
    \fl{e} Perturbative expansion of first order and second order derivatives, highlighting that the higher order derivative does not vanish at the same $\phi$ value as the first order.
    \fl{f} Measurement of \var{\tilde{S}}{1} when detuning QDR, with the settings in \cref{fig:Fig4}c, in order to compare with the predicted spectrum (d). 
    Repeated measurements for a range of \var{B}{x} values is shown in \cref{fig:S9_detuning_vs_phase}.}
    \label{fig:Fig5}
\end{figure}
where $\Delta^{(1,2)}=\Delta^{(2,3)}$ and $t^{(1,2)}=t^{(2,3)}$.
Along this line (\cref{fig:Fig5}a), the Hamiltonian in Eq. \eqref{eq:3_site_effective} commutes with the total fermion parity operator $P_F = \prod_{i,\sigma} (1-2n_{i,\sigma})$, and a $\mathbb{Z}_3$ parity operator $P_3 = \exp\left[i\frac{2\pi}{3}\sum_i(n_{i,\uparrow}+2n_{i,\downarrow})\right]$ \cite{teixeria2022}.
The emergence of this commutator relation can be understood from the point of chirality~\cite{Alicea2016}.
The 27 levels contained in the low-energy subspace split into 7 distinct manifolds: 3 triply-degenerate levels with $E<0$, 1 nine-fold degenerate manifold at $E=0$, and another 3 triply-degenerate manifold with $E>0$ (see Methods).\newline\newline
Ideally, a point can be found where the strongly degenerate spectrum again gives rise to Majorana Kramers Pairs or $\mathbb{Z}_3$ parafermions, as in~\cite{bozkurt2025}.
To assess this, we numerically calculate the linear derivatives of the energy splitting between odd and even ground states with respect to the individual dot chemical potentials, $\partial_{\mu_i}(E_{even,0}-E_{odd,0})$ (\cref{fig:Fig5}b).
Doing so for all $\varphi$, we find that the ground state degeneracy is unaffected by changes of the chemical potential of the central dot, $\mu_2$, in linear order.
In contrast, the degeneracy splits linearly when varying the chemical potential of the outer dots, $\mu_1, \mu_3$, except for two points where the coupling in linear order of perturbation theory vanishes.
Numerically these phases are found to be at $\varphi^{*} \in \{1.86, 4.42\}$. 
To verify this, we numerically calculate a perturbative expansion for the ground state manifold when perturbed by any $\mu_i (n_{i\ua}+n_{i\da})$\cite{araya2025}, i.e. local perturbations to the chemical potential of one of the QDs.
Since the perturbations neither break parity nor time-reversal symmetry, we can understand the system's response by considering the ratio $\delta E_{\mathrm{even}}/\delta E_{\mathrm{odd}}$ of the energy shifts $E_i\rightarrow E_{i}^{(0)}+\delta E_{i}$ induced within each parity sector.
To linear order in $\mu_i$, we indeed recover the spectral insensitivity to changes as $\delta E_{\mathrm{even}}(\mu_i) = \delta E_{\mathrm{odd}}(\mu_i)$ at $\varphi^{*}_i$.
Expanding to second order in $\mu_i$, we however find that the splitting at $\varphi^*_i$ differs and therefore breaks the degeneracy (\cref{fig:Fig5}e).
Consequently, we will not be able to find Majorana Kramers pair or $\mathbb{Z}_3$ parafermions operators that are robust against local perturbations as previously \cite{bozkurt2025} for two sites.
We note that this is in line with previous conclusions for strong degeneracies in chiral spin chains~\cite{Fendley2012}.
To conclude, we consider the conductance signatures at two points along the triply degenerate line (\cref{fig:Fig5}c,d).
While the perturbative results suggest a splitting of the levels with the chemical potential, we find in conductance that this splitting can become very small and not possible to resolve due to broadening from temperature effects.
Experimentally, we indeed observe that a phase value can be found where a seemingly robust zero-bias conductance peak appears (\cref{fig:Fig5}f).
We want to highlight that this behaviour is not to be generally expected, but only along this fine-tuned line of parameters.
\section*{Conclusions}
We have studied a superconducting extension of the Fermi-Hubbard model in short arrays of semiconductor-superconductor hybrid quantum dot arrays.
In the absence of spin-polarizing magnetic fields, the on-site Coulomb repulsion is the largest energy scale of the system.
Tuning the normal and superconducting couplings allows to deliberately tune the system to host degenerate spectra.
For both two and three electronic sites we explored the signatures of such degeneracies.
In the case of two sites, we found close agreement with the predictions put forward in~\cite{bozkurt2025}, suggesting robust Majorana Kramers pairs~\cite{Wright2013, OBrien2015} or $\mathbb{Z}_3$-parafermions at specific parameter configurations.
We identify these configurations by experimentally finding robust zero-bias peaks.
Notably, such excitations can have applications in topological quantum computing~\cite{Alicea2016,Gao2016,Schrade2022, Pan2025}, but were typically so far only predicted for more intricate physical platforms~\cite{Wong2012,FZhang2013,Keselman2013,Klinovaja2014,Klinovaja2014b,Haim2014,XJLiu2014,Schrade2015,Thakurathi2018}.
Whether our platform, hosting highly non-local Kramers pairs, has potential use in this context remains subject of further research.
Upon extending the system to three sites, we do find signatures of degeneracies by sweeping the system's larger parameter space.
A theoretical analysis shows these to be consistent with strongly degenerate spectra along fine-tuned configurations of the system.
In contrast to the two-site system, these degeneracies are however not robust but predicted to split quadratically.
This we interpret as being consistent with previous predictions that strong degeneracies in quasi one-dimensional systems are generally not robust against perturbations~\cite{jermyn2014,Alicea2016}.
Based on our observation, we expect hybrid superconductor - quantum dot systems to allow for analysis of the stability of strong degeneracies through quantum simulation for systems beyond computational feasibility.
Furthermore, extending the system to two dimensions, we foresee the platform to be useful in understanding strongly correlated phenomena when subjected to superconducting correlations.
\section{Acknowledgements}
We thank C. Thomas, D. Xiao and M. J. Manfra for the provision of the 2DEG materials. 
We thank O.W.B.~Benningshof, T.~Orton and J.D.~Mensingh for technical assistance with the cryogenic electronics.
We thank M. Burrello for his comments on the fractionalized modes and F. Zatelli for valuable input on the manuscript.
The research at Delft was supported by the Dutch National Science Foundation (NWO),  Microsoft Corporation Station Q and a grant from Top consortium for Knowledge and Innovation program (TKI). 
S.G. and M.W acknowledge financial support from the Horizon Europe Framework Program of the European Commission through the European Innovation Council Pathfinder grant no. 101115315 (QuKiT).
S.M. acknowledges funding from the Dutch Organization for Scientific Research (NWO) through OCENW.GROOT.2019.004.

\section{Author contributions}
Q.W., I.K. and S.L.D.tH fabricated the devices.
Y.Z. and I.K. fabricated the resonator circuits.
C.G.P. and I.K. designed the measurement set-up.
Measurements were performed by S.L.D.tH.
Theoretical analysis was performed by S.M., S.L.D.tH., A.M.B. and M.W.
The manuscript was written by S.M., S.L.D.tH., M.W. and S.G., with input from all co-authors. 
S.G. supervised the experimental work in Delft.

\section{Data availability}
All raw data obtained in relation to this manuscript, the code to generate the theoretical results and the scripts to reproduce the figures from the raw data are available on Zenodo~\cite{ZenodoDataRepo}.

\bibliography{ref} 

\begin{thebibliography}{63}%
\makeatletter
\providecommand \@ifxundefined [1]{%
 \@ifx{#1\undefined}
}%
\providecommand \@ifnum [1]{%
 \ifnum #1\expandafter \@firstoftwo
 \else \expandafter \@secondoftwo
 \fi
}%
\providecommand \@ifx [1]{%
 \ifx #1\expandafter \@firstoftwo
 \else \expandafter \@secondoftwo
 \fi
}%
\providecommand \natexlab [1]{#1}%
\providecommand \enquote  [1]{``#1''}%
\providecommand \bibnamefont  [1]{#1}%
\providecommand \bibfnamefont [1]{#1}%
\providecommand \citenamefont [1]{#1}%
\providecommand \href@noop [0]{\@secondoftwo}%
\providecommand \href [0]{\begingroup \@sanitize@url \@href}%
\providecommand \@href[1]{\@@startlink{#1}\@@href}%
\providecommand \@@href[1]{\endgroup#1\@@endlink}%
\providecommand \@sanitize@url [0]{\catcode `\\12\catcode `\$12\catcode `\&12\catcode `\#12\catcode `\^12\catcode `\_12\catcode `\%12\relax}%
\providecommand \@@startlink[1]{}%
\providecommand \@@endlink[0]{}%
\providecommand \url  [0]{\begingroup\@sanitize@url \@url }%
\providecommand \@url [1]{\endgroup\@href {#1}{\urlprefix }}%
\providecommand \urlprefix  [0]{URL }%
\providecommand \Eprint [0]{\href }%
\providecommand \doibase [0]{https://doi.org/}%
\providecommand \selectlanguage [0]{\@gobble}%
\providecommand \bibinfo  [0]{\@secondoftwo}%
\providecommand \bibfield  [0]{\@secondoftwo}%
\providecommand \translation [1]{[#1]}%
\providecommand \BibitemOpen [0]{}%
\providecommand \bibitemStop [0]{}%
\providecommand \bibitemNoStop [0]{.\EOS\space}%
\providecommand \EOS [0]{\spacefactor3000\relax}%
\providecommand \BibitemShut  [1]{\csname bibitem#1\endcsname}%
\let\auto@bib@innerbib\@empty
\bibitem [{\citenamefont {Hubbard}\ and\ \citenamefont {Flowers}(1963)}]{hubbard_model}%
  \BibitemOpen
  \bibfield  {author} {\bibinfo {author} {\bibfnamefont {J.}~\bibnamefont {Hubbard}}\ and\ \bibinfo {author} {\bibfnamefont {B.~H.}\ \bibnamefont {Flowers}},\ }\bibfield  {title} {\bibinfo {title} {Electron correlations in narrow energy bands},\ }\href {https://doi.org/10.1098/rspa.1963.0204} {\bibfield  {journal} {\bibinfo  {journal} {Proceedings of the Royal Society of London. Series A. Mathematical and Physical Sciences}\ }\textbf {\bibinfo {volume} {276}},\ \bibinfo {pages} {238} (\bibinfo {year} {1963})}\BibitemShut {NoStop}%
\bibitem [{\citenamefont {Arovas}\ \emph {et~al.}(2022)\citenamefont {Arovas}, \citenamefont {Berg}, \citenamefont {Kivelson},\ and\ \citenamefont {Raghu}}]{hubbard_model_review}%
  \BibitemOpen
  \bibfield  {author} {\bibinfo {author} {\bibfnamefont {D.~P.}\ \bibnamefont {Arovas}}, \bibinfo {author} {\bibfnamefont {E.}~\bibnamefont {Berg}}, \bibinfo {author} {\bibfnamefont {S.~A.}\ \bibnamefont {Kivelson}},\ and\ \bibinfo {author} {\bibfnamefont {S.}~\bibnamefont {Raghu}},\ }\bibfield  {title} {\bibinfo {title} {The hubbard model},\ }\href {https://doi.org/https://doi.org/10.1146/annurev-conmatphys-031620-102024} {\bibfield  {journal} {\bibinfo  {journal} {Annual Review of Condensed Matter Physics}\ }\textbf {\bibinfo {volume} {13}},\ \bibinfo {pages} {239} (\bibinfo {year} {2022})}\BibitemShut {NoStop}%
\bibitem [{\citenamefont {Lee}\ \emph {et~al.}(2006)\citenamefont {Lee}, \citenamefont {Nagaosa},\ and\ \citenamefont {Wen}}]{lee_doping_2006}%
  \BibitemOpen
  \bibfield  {author} {\bibinfo {author} {\bibfnamefont {P.~A.}\ \bibnamefont {Lee}}, \bibinfo {author} {\bibfnamefont {N.}~\bibnamefont {Nagaosa}},\ and\ \bibinfo {author} {\bibfnamefont {X.-G.}\ \bibnamefont {Wen}},\ }\bibfield  {title} {\bibinfo {title} {Doping a {Mott} insulator: {Physics} of high-temperature superconductivity},\ }\href {https://doi.org/10.1103/RevModPhys.78.17} {\bibfield  {journal} {\bibinfo  {journal} {Reviews of Modern Physics}\ }\textbf {\bibinfo {volume} {78}},\ \bibinfo {pages} {17} (\bibinfo {year} {2006})},\ \bibinfo {note} {publisher: American Physical Society}\BibitemShut {NoStop}%
\bibitem [{\citenamefont {Esslinger}(2010)}]{esslinger_fermi-hubbard_2010}%
  \BibitemOpen
  \bibfield  {author} {\bibinfo {author} {\bibfnamefont {T.}~\bibnamefont {Esslinger}},\ }\bibfield  {title} {\bibinfo {title} {Fermi-{Hubbard} {Physics} with {Atoms} in an {Optical} {Lattice}},\ }\href {https://doi.org/10.1146/annurev-conmatphys-070909-104059} {\bibfield  {journal} {\bibinfo  {journal} {Annual Review of Condensed Matter Physics}\ }\textbf {\bibinfo {volume} {1}},\ \bibinfo {pages} {129} (\bibinfo {year} {2010})},\ \bibinfo {note} {publisher: Annual Reviews}\BibitemShut {NoStop}%
\bibitem [{\citenamefont {Johanning}\ \emph {et~al.}(2009)\citenamefont {Johanning}, \citenamefont {Varón},\ and\ \citenamefont {Wunderlich}}]{johanning_quantum_2009}%
  \BibitemOpen
  \bibfield  {author} {\bibinfo {author} {\bibfnamefont {M.}~\bibnamefont {Johanning}}, \bibinfo {author} {\bibfnamefont {A.~F.}\ \bibnamefont {Varón}},\ and\ \bibinfo {author} {\bibfnamefont {C.}~\bibnamefont {Wunderlich}},\ }\bibfield  {title} {\bibinfo {title} {Quantum simulations with cold trapped ions},\ }\href {https://doi.org/10.1088/0953-4075/42/15/154009} {\bibfield  {journal} {\bibinfo  {journal} {Journal of Physics B: Atomic, Molecular and Optical Physics}\ }\textbf {\bibinfo {volume} {42}},\ \bibinfo {pages} {154009} (\bibinfo {year} {2009})}\BibitemShut {NoStop}%
\bibitem [{\citenamefont {Schmidt}\ and\ \citenamefont {Koch}(2013)}]{schmidt_circuit_2013}%
  \BibitemOpen
  \bibfield  {author} {\bibinfo {author} {\bibfnamefont {S.}~\bibnamefont {Schmidt}}\ and\ \bibinfo {author} {\bibfnamefont {J.}~\bibnamefont {Koch}},\ }\bibfield  {title} {\bibinfo {title} {Circuit {QED} lattices: {Towards} quantum simulation with superconducting circuits},\ }\href {https://doi.org/10.1002/andp.201200261} {\bibfield  {journal} {\bibinfo  {journal} {Annalen der Physik}\ }\textbf {\bibinfo {volume} {525}},\ \bibinfo {pages} {395} (\bibinfo {year} {2013})}\BibitemShut {NoStop}%
\bibitem [{\citenamefont {Houck}\ \emph {et~al.}(2012)\citenamefont {Houck}, \citenamefont {Türeci},\ and\ \citenamefont {Koch}}]{houck_-chip_2012}%
  \BibitemOpen
  \bibfield  {author} {\bibinfo {author} {\bibfnamefont {A.~A.}\ \bibnamefont {Houck}}, \bibinfo {author} {\bibfnamefont {H.~E.}\ \bibnamefont {Türeci}},\ and\ \bibinfo {author} {\bibfnamefont {J.}~\bibnamefont {Koch}},\ }\bibfield  {title} {\bibinfo {title} {On-chip quantum simulation with superconducting circuits},\ }\href {https://doi.org/10.1038/nphys2251} {\bibfield  {journal} {\bibinfo  {journal} {Nature Physics}\ }\textbf {\bibinfo {volume} {8}},\ \bibinfo {pages} {292} (\bibinfo {year} {2012})},\ \bibinfo {note} {publisher: Nature Publishing Group}\BibitemShut {NoStop}%
\bibitem [{\citenamefont {Yang}\ \emph {et~al.}(2011)\citenamefont {Yang}, \citenamefont {Wang},\ and\ \citenamefont {Das~Sarma}}]{yang_generic_2011}%
  \BibitemOpen
  \bibfield  {author} {\bibinfo {author} {\bibfnamefont {S.}~\bibnamefont {Yang}}, \bibinfo {author} {\bibfnamefont {X.}~\bibnamefont {Wang}},\ and\ \bibinfo {author} {\bibfnamefont {S.}~\bibnamefont {Das~Sarma}},\ }\bibfield  {title} {\bibinfo {title} {Generic {Hubbard} model description of semiconductor quantum-dot spin qubits},\ }\href {https://doi.org/10.1103/PhysRevB.83.161301} {\bibfield  {journal} {\bibinfo  {journal} {Physical Review B}\ }\textbf {\bibinfo {volume} {83}},\ \bibinfo {pages} {161301} (\bibinfo {year} {2011})},\ \bibinfo {note} {publisher: American Physical Society}\BibitemShut {NoStop}%
\bibitem [{\citenamefont {Hensgens}\ \emph {et~al.}(2017)\citenamefont {Hensgens}, \citenamefont {Fujita}, \citenamefont {Janssen}, \citenamefont {Li}, \citenamefont {Van~Diepen}, \citenamefont {Reichl}, \citenamefont {Wegscheider}, \citenamefont {Das~Sarma},\ and\ \citenamefont {Vandersypen}}]{hensgens_quantum_2017}%
  \BibitemOpen
  \bibfield  {author} {\bibinfo {author} {\bibfnamefont {T.}~\bibnamefont {Hensgens}}, \bibinfo {author} {\bibfnamefont {T.}~\bibnamefont {Fujita}}, \bibinfo {author} {\bibfnamefont {L.}~\bibnamefont {Janssen}}, \bibinfo {author} {\bibfnamefont {X.}~\bibnamefont {Li}}, \bibinfo {author} {\bibfnamefont {C.~J.}\ \bibnamefont {Van~Diepen}}, \bibinfo {author} {\bibfnamefont {C.}~\bibnamefont {Reichl}}, \bibinfo {author} {\bibfnamefont {W.}~\bibnamefont {Wegscheider}}, \bibinfo {author} {\bibfnamefont {S.}~\bibnamefont {Das~Sarma}},\ and\ \bibinfo {author} {\bibfnamefont {L.~M.~K.}\ \bibnamefont {Vandersypen}},\ }\bibfield  {title} {\bibinfo {title} {Quantum simulation of a {Fermi}–{Hubbard} model using a semiconductor quantum dot array},\ }\href {https://doi.org/10.1038/nature23022} {\bibfield  {journal} {\bibinfo  {journal} {Nature}\ }\textbf {\bibinfo {volume} {548}},\ \bibinfo {pages} {70} (\bibinfo {year} {2017})},\ \bibinfo {note} {publisher: Nature Publishing Group}\BibitemShut {NoStop}%
\bibitem [{\citenamefont {Byrnes}\ \emph {et~al.}(2008)\citenamefont {Byrnes}, \citenamefont {Kim}, \citenamefont {Kusudo},\ and\ \citenamefont {Yamamoto}}]{byrnes_quantum_2008}%
  \BibitemOpen
  \bibfield  {author} {\bibinfo {author} {\bibfnamefont {T.}~\bibnamefont {Byrnes}}, \bibinfo {author} {\bibfnamefont {N.~Y.}\ \bibnamefont {Kim}}, \bibinfo {author} {\bibfnamefont {K.}~\bibnamefont {Kusudo}},\ and\ \bibinfo {author} {\bibfnamefont {Y.}~\bibnamefont {Yamamoto}},\ }\bibfield  {title} {\bibinfo {title} {Quantum simulation of {Fermi}-{Hubbard} models in semiconductor quantum-dot arrays},\ }\href {https://doi.org/10.1103/PhysRevB.78.075320} {\bibfield  {journal} {\bibinfo  {journal} {Physical Review B}\ }\textbf {\bibinfo {volume} {78}},\ \bibinfo {pages} {075320} (\bibinfo {year} {2008})},\ \bibinfo {note} {publisher: American Physical Society}\BibitemShut {NoStop}%
\bibitem [{\citenamefont {Kitaev}(2001)}]{Kitaev2001}%
  \BibitemOpen
  \bibfield  {author} {\bibinfo {author} {\bibfnamefont {A.~Y.}\ \bibnamefont {Kitaev}},\ }\bibfield  {title} {\bibinfo {title} {Unpaired {Majorana} fermions in quantum wires},\ }\href {https://dx.doi.org/10.1070/1063-7869/44/10S/S29} {\bibfield  {journal} {\bibinfo  {journal} {Physics-Usphekhi}\ }\textbf {\bibinfo {volume} {44}} (\bibinfo {year} {2001})}\BibitemShut {NoStop}%
\bibitem [{\citenamefont {Sau}\ and\ \citenamefont {Sarma}(2012)}]{Sau2012}%
  \BibitemOpen
  \bibfield  {author} {\bibinfo {author} {\bibfnamefont {J.~D.}\ \bibnamefont {Sau}}\ and\ \bibinfo {author} {\bibfnamefont {S.~D.}\ \bibnamefont {Sarma}},\ }\bibfield  {title} {\bibinfo {title} {Realizing a robust practical {Majorana} chain in a quantum-dot-superconductor linear array},\ }\href {https://doi.org/10.1038/ncomms1966} {\bibfield  {journal} {\bibinfo  {journal} {Nature Communications}\ }\textbf {\bibinfo {volume} {3}},\ \bibinfo {pages} {964} (\bibinfo {year} {2012})}\BibitemShut {NoStop}%
\bibitem [{\citenamefont {Leijnse}\ and\ \citenamefont {Flensberg}(2012)}]{Leijnse2012}%
  \BibitemOpen
  \bibfield  {author} {\bibinfo {author} {\bibfnamefont {M.}~\bibnamefont {Leijnse}}\ and\ \bibinfo {author} {\bibfnamefont {K.}~\bibnamefont {Flensberg}},\ }\bibfield  {title} {\bibinfo {title} {Parity qubits and poor man's {Majorana} bound states in double quantum dots},\ }\href {https://doi.org/10.1103/PhysRevB.86.134528} {\bibfield  {journal} {\bibinfo  {journal} {Phys. Rev. B}\ }\textbf {\bibinfo {volume} {86}},\ \bibinfo {pages} {134528} (\bibinfo {year} {2012})}\BibitemShut {NoStop}%
\bibitem [{\citenamefont {Fulga}\ \emph {et~al.}(2013)\citenamefont {Fulga}, \citenamefont {Haim}, \citenamefont {Akhmerov},\ and\ \citenamefont {Oreg}}]{Fulga2013}%
  \BibitemOpen
  \bibfield  {author} {\bibinfo {author} {\bibfnamefont {I.~C.}\ \bibnamefont {Fulga}}, \bibinfo {author} {\bibfnamefont {A.}~\bibnamefont {Haim}}, \bibinfo {author} {\bibfnamefont {A.~R.}\ \bibnamefont {Akhmerov}},\ and\ \bibinfo {author} {\bibfnamefont {Y.}~\bibnamefont {Oreg}},\ }\bibfield  {title} {\bibinfo {title} {Adaptive tuning of {Majorana} fermions in a quantum dot chain},\ }\href {https://doi.org/10.1088/1367-2630/15/4/045020} {\bibfield  {journal} {\bibinfo  {journal} {New Journal of Physics}\ }\textbf {\bibinfo {volume} {15}},\ \bibinfo {pages} {045020} (\bibinfo {year} {2013})}\BibitemShut {NoStop}%
\bibitem [{\citenamefont {Dvir}\ \emph {et~al.}(2023)\citenamefont {Dvir}, \citenamefont {Wang}, \citenamefont {van Loo}, \citenamefont {Liu}, \citenamefont {Mazur}, \citenamefont {Bordin}, \citenamefont {ten Haaf}, \citenamefont {Wang}, \citenamefont {van Driel}, \citenamefont {Zatelli}, \citenamefont {Li}, \citenamefont {Malinowski}, \citenamefont {Gazibegovic}, \citenamefont {Badawy}, \citenamefont {Bakkers}, \citenamefont {Wimmer},\ and\ \citenamefont {Kouwenhoven}}]{Dvir2023}%
  \BibitemOpen
  \bibfield  {author} {\bibinfo {author} {\bibfnamefont {T.}~\bibnamefont {Dvir}}, \bibinfo {author} {\bibfnamefont {G.}~\bibnamefont {Wang}}, \bibinfo {author} {\bibfnamefont {N.}~\bibnamefont {van Loo}}, \bibinfo {author} {\bibfnamefont {C.-X.}\ \bibnamefont {Liu}}, \bibinfo {author} {\bibfnamefont {G.~P.}\ \bibnamefont {Mazur}}, \bibinfo {author} {\bibfnamefont {A.}~\bibnamefont {Bordin}}, \bibinfo {author} {\bibfnamefont {S.~L.~D.}\ \bibnamefont {ten Haaf}}, \bibinfo {author} {\bibfnamefont {J.-Y.}\ \bibnamefont {Wang}}, \bibinfo {author} {\bibfnamefont {D.}~\bibnamefont {van Driel}}, \bibinfo {author} {\bibfnamefont {F.}~\bibnamefont {Zatelli}}, \bibinfo {author} {\bibfnamefont {X.}~\bibnamefont {Li}}, \bibinfo {author} {\bibfnamefont {F.~K.}\ \bibnamefont {Malinowski}}, \bibinfo {author} {\bibfnamefont {S.}~\bibnamefont {Gazibegovic}}, \bibinfo {author} {\bibfnamefont {G.}~\bibnamefont {Badawy}}, \bibinfo {author} {\bibfnamefont {E.~P. A.~M.}\ \bibnamefont {Bakkers}}, \bibinfo {author} {\bibfnamefont
  {M.}~\bibnamefont {Wimmer}},\ and\ \bibinfo {author} {\bibfnamefont {L.~P.}\ \bibnamefont {Kouwenhoven}},\ }\bibfield  {title} {\bibinfo {title} {Realization of a minimal {Kitaev} chain in coupled quantum dots},\ }\href {https://doi.org/10.1038/s41586-022-05585-1} {\bibfield  {journal} {\bibinfo  {journal} {Nature}\ }\textbf {\bibinfo {volume} {614}},\ \bibinfo {pages} {445} (\bibinfo {year} {2023})}\BibitemShut {NoStop}%
\bibitem [{\citenamefont {ten Haaf}\ \emph {et~al.}(2024)\citenamefont {ten Haaf}, \citenamefont {Wang}, \citenamefont {Bozkurt}, \citenamefont {Liu}, \citenamefont {Kulesh}, \citenamefont {Kim}, \citenamefont {Xiao}, \citenamefont {Thomas}, \citenamefont {Manfra}, \citenamefont {Dvir}, \citenamefont {Wimmer},\ and\ \citenamefont {Goswami}}]{tenHaaf2024}%
  \BibitemOpen
  \bibfield  {author} {\bibinfo {author} {\bibfnamefont {S.~L.~D.}\ \bibnamefont {ten Haaf}}, \bibinfo {author} {\bibfnamefont {Q.}~\bibnamefont {Wang}}, \bibinfo {author} {\bibfnamefont {A.~M.}\ \bibnamefont {Bozkurt}}, \bibinfo {author} {\bibfnamefont {C.-X.}\ \bibnamefont {Liu}}, \bibinfo {author} {\bibfnamefont {I.}~\bibnamefont {Kulesh}}, \bibinfo {author} {\bibfnamefont {P.}~\bibnamefont {Kim}}, \bibinfo {author} {\bibfnamefont {D.}~\bibnamefont {Xiao}}, \bibinfo {author} {\bibfnamefont {C.}~\bibnamefont {Thomas}}, \bibinfo {author} {\bibfnamefont {M.~J.}\ \bibnamefont {Manfra}}, \bibinfo {author} {\bibfnamefont {T.}~\bibnamefont {Dvir}}, \bibinfo {author} {\bibfnamefont {M.}~\bibnamefont {Wimmer}},\ and\ \bibinfo {author} {\bibfnamefont {S.}~\bibnamefont {Goswami}},\ }\bibfield  {title} {\bibinfo {title} {A two-site {Kitaev} chain in a two-dimensional electron gas},\ }\href {https://doi.org/10.1038/s41586-024-07434-9} {\bibfield  {journal} {\bibinfo  {journal} {Nature}\ }\textbf {\bibinfo {volume}
  {630}},\ \bibinfo {pages} {329} (\bibinfo {year} {2024})}\BibitemShut {NoStop}%
\bibitem [{\citenamefont {Bordin}\ \emph {et~al.}(2025{\natexlab{a}})\citenamefont {Bordin}, \citenamefont {Liu}, \citenamefont {Dvir}, \citenamefont {Zatelli}, \citenamefont {ten Haaf}, \citenamefont {van Driel}, \citenamefont {Wang}, \citenamefont {van Loo}, \citenamefont {Zhang}, \citenamefont {Wolff}, \citenamefont {Van~Caekenberghe}, \citenamefont {Badawy}, \citenamefont {Gazibegovic}, \citenamefont {Bakkers}, \citenamefont {Wimmer}, \citenamefont {Kouwenhoven},\ and\ \citenamefont {Mazur}}]{bordin2024}%
  \BibitemOpen
  \bibfield  {author} {\bibinfo {author} {\bibfnamefont {A.}~\bibnamefont {Bordin}}, \bibinfo {author} {\bibfnamefont {C.-X.}\ \bibnamefont {Liu}}, \bibinfo {author} {\bibfnamefont {T.}~\bibnamefont {Dvir}}, \bibinfo {author} {\bibfnamefont {F.}~\bibnamefont {Zatelli}}, \bibinfo {author} {\bibfnamefont {S.~L.~D.}\ \bibnamefont {ten Haaf}}, \bibinfo {author} {\bibfnamefont {D.}~\bibnamefont {van Driel}}, \bibinfo {author} {\bibfnamefont {G.}~\bibnamefont {Wang}}, \bibinfo {author} {\bibfnamefont {N.}~\bibnamefont {van Loo}}, \bibinfo {author} {\bibfnamefont {Y.}~\bibnamefont {Zhang}}, \bibinfo {author} {\bibfnamefont {J.~C.}\ \bibnamefont {Wolff}}, \bibinfo {author} {\bibfnamefont {T.}~\bibnamefont {Van~Caekenberghe}}, \bibinfo {author} {\bibfnamefont {G.}~\bibnamefont {Badawy}}, \bibinfo {author} {\bibfnamefont {S.}~\bibnamefont {Gazibegovic}}, \bibinfo {author} {\bibfnamefont {E.~P. A.~M.}\ \bibnamefont {Bakkers}}, \bibinfo {author} {\bibfnamefont {M.}~\bibnamefont {Wimmer}}, \bibinfo {author} {\bibfnamefont
  {L.~P.}\ \bibnamefont {Kouwenhoven}},\ and\ \bibinfo {author} {\bibfnamefont {G.~P.}\ \bibnamefont {Mazur}},\ }\bibfield  {title} {\bibinfo {title} {Enhanced {Majorana} stability in a three-site {Kitaev} chain},\ }\href {https://doi.org/10.1038/s41565-025-01894-4} {\bibfield  {journal} {\bibinfo  {journal} {Nature Nanotechnology}\ }\textbf {\bibinfo {volume} {20}},\ \bibinfo {pages} {726} (\bibinfo {year} {2025}{\natexlab{a}})}\BibitemShut {NoStop}%
\bibitem [{\citenamefont {ten Haaf}\ \emph {et~al.}(2025)\citenamefont {ten Haaf}, \citenamefont {Zhang}, \citenamefont {Wang}, \citenamefont {Bordin}, \citenamefont {Liu}, \citenamefont {Kulesh}, \citenamefont {Sietses}, \citenamefont {Prosko}, \citenamefont {Xiao}, \citenamefont {Thomas}, \citenamefont {Manfra}, \citenamefont {Wimmer},\ and\ \citenamefont {Goswami}}]{haaf2025}%
  \BibitemOpen
  \bibfield  {author} {\bibinfo {author} {\bibfnamefont {S.~L.~D.}\ \bibnamefont {ten Haaf}}, \bibinfo {author} {\bibfnamefont {Y.}~\bibnamefont {Zhang}}, \bibinfo {author} {\bibfnamefont {Q.}~\bibnamefont {Wang}}, \bibinfo {author} {\bibfnamefont {A.}~\bibnamefont {Bordin}}, \bibinfo {author} {\bibfnamefont {C.-X.}\ \bibnamefont {Liu}}, \bibinfo {author} {\bibfnamefont {I.}~\bibnamefont {Kulesh}}, \bibinfo {author} {\bibfnamefont {V.~P.~M.}\ \bibnamefont {Sietses}}, \bibinfo {author} {\bibfnamefont {C.~G.}\ \bibnamefont {Prosko}}, \bibinfo {author} {\bibfnamefont {D.}~\bibnamefont {Xiao}}, \bibinfo {author} {\bibfnamefont {C.}~\bibnamefont {Thomas}}, \bibinfo {author} {\bibfnamefont {M.~J.}\ \bibnamefont {Manfra}}, \bibinfo {author} {\bibfnamefont {M.}~\bibnamefont {Wimmer}},\ and\ \bibinfo {author} {\bibfnamefont {S.}~\bibnamefont {Goswami}},\ }\bibfield  {title} {\bibinfo {title} {Observation of edge and bulk states in a three-site {Kitaev} chain},\ }\href {https://doi.org/10.1038/s41586-025-08892-5}
  {\bibfield  {journal} {\bibinfo  {journal} {Nature}\ }\textbf {\bibinfo {volume} {641}},\ \bibinfo {pages} {890} (\bibinfo {year} {2025})}\BibitemShut {NoStop}%
\bibitem [{\citenamefont {Liu}\ \emph {et~al.}(2022)\citenamefont {Liu}, \citenamefont {Wang}, \citenamefont {Dvir},\ and\ \citenamefont {Wimmer}}]{CXLiu2022}%
  \BibitemOpen
  \bibfield  {author} {\bibinfo {author} {\bibfnamefont {C.-X.}\ \bibnamefont {Liu}}, \bibinfo {author} {\bibfnamefont {G.}~\bibnamefont {Wang}}, \bibinfo {author} {\bibfnamefont {T.}~\bibnamefont {Dvir}},\ and\ \bibinfo {author} {\bibfnamefont {M.}~\bibnamefont {Wimmer}},\ }\bibfield  {title} {\bibinfo {title} {Tunable superconducting coupling of quantum dots via {Andreev} bound states in semiconductor-superconductor nanowires},\ }\href {https://doi.org/10.1103/PhysRevLett.129.267701} {\bibfield  {journal} {\bibinfo  {journal} {Phys. Rev. Lett.}\ }\textbf {\bibinfo {volume} {129}},\ \bibinfo {pages} {267701} (\bibinfo {year} {2022})}\BibitemShut {NoStop}%
\bibitem [{\citenamefont {Tsintzis}\ \emph {et~al.}(2022)\citenamefont {Tsintzis}, \citenamefont {Souto},\ and\ \citenamefont {Leijnse}}]{Tsintzis20222}%
  \BibitemOpen
  \bibfield  {author} {\bibinfo {author} {\bibfnamefont {A.}~\bibnamefont {Tsintzis}}, \bibinfo {author} {\bibfnamefont {R.~S.}\ \bibnamefont {Souto}},\ and\ \bibinfo {author} {\bibfnamefont {M.}~\bibnamefont {Leijnse}},\ }\bibfield  {title} {\bibinfo {title} {Creating and detecting poor man's {Majorana} bound states in interacting quantum dots},\ }\href {https://doi.org/10.1103/PhysRevB.106.L201404} {\bibfield  {journal} {\bibinfo  {journal} {Phys. Rev. B}\ }\textbf {\bibinfo {volume} {106}},\ \bibinfo {pages} {L201404} (\bibinfo {year} {2022})}\BibitemShut {NoStop}%
\bibitem [{\citenamefont {Zatelli}\ \emph {et~al.}(2024)\citenamefont {Zatelli}, \citenamefont {van Driel}, \citenamefont {Xu}, \citenamefont {Wang}, \citenamefont {Liu}, \citenamefont {Bordin}, \citenamefont {Roovers}, \citenamefont {Mazur}, \citenamefont {van Loo}, \citenamefont {Wolff}, \citenamefont {Bozkurt}, \citenamefont {Badawy}, \citenamefont {Gazibegovic}, \citenamefont {Bakkers}, \citenamefont {Wimmer}, \citenamefont {Kouwenhoven},\ and\ \citenamefont {Dvir}}]{Zatelli2024}%
  \BibitemOpen
  \bibfield  {author} {\bibinfo {author} {\bibfnamefont {F.}~\bibnamefont {Zatelli}}, \bibinfo {author} {\bibfnamefont {D.}~\bibnamefont {van Driel}}, \bibinfo {author} {\bibfnamefont {D.}~\bibnamefont {Xu}}, \bibinfo {author} {\bibfnamefont {G.}~\bibnamefont {Wang}}, \bibinfo {author} {\bibfnamefont {C.-X.}\ \bibnamefont {Liu}}, \bibinfo {author} {\bibfnamefont {A.}~\bibnamefont {Bordin}}, \bibinfo {author} {\bibfnamefont {B.}~\bibnamefont {Roovers}}, \bibinfo {author} {\bibfnamefont {G.~P.}\ \bibnamefont {Mazur}}, \bibinfo {author} {\bibfnamefont {N.}~\bibnamefont {van Loo}}, \bibinfo {author} {\bibfnamefont {J.~C.}\ \bibnamefont {Wolff}}, \bibinfo {author} {\bibfnamefont {A.~M.}\ \bibnamefont {Bozkurt}}, \bibinfo {author} {\bibfnamefont {G.}~\bibnamefont {Badawy}}, \bibinfo {author} {\bibfnamefont {S.}~\bibnamefont {Gazibegovic}}, \bibinfo {author} {\bibfnamefont {E.~P. A.~M.}\ \bibnamefont {Bakkers}}, \bibinfo {author} {\bibfnamefont {M.}~\bibnamefont {Wimmer}}, \bibinfo {author} {\bibfnamefont {L.~P.}\
  \bibnamefont {Kouwenhoven}},\ and\ \bibinfo {author} {\bibfnamefont {T.}~\bibnamefont {Dvir}},\ }\bibfield  {title} {\bibinfo {title} {Robust poor man's {Majorana} zero modes using {Yu}-{Shiba}-{Rusinov} states},\ }\href {https://doi.org/10.1038/s41467-024-52066-2} {\bibfield  {journal} {\bibinfo  {journal} {Nature Communications}\ }\textbf {\bibinfo {volume} {15}},\ \bibinfo {pages} {7933} (\bibinfo {year} {2024})}\BibitemShut {NoStop}%
\bibitem [{\citenamefont {Bozkurt}\ \emph {et~al.}(2025)\citenamefont {Bozkurt}, \citenamefont {Miles}, \citenamefont {ten Haaf}, \citenamefont {Liu}, \citenamefont {Hassler},\ and\ \citenamefont {Wimmer}}]{bozkurt2025}%
  \BibitemOpen
  \bibfield  {author} {\bibinfo {author} {\bibfnamefont {A.~M.}\ \bibnamefont {Bozkurt}}, \bibinfo {author} {\bibfnamefont {S.}~\bibnamefont {Miles}}, \bibinfo {author} {\bibfnamefont {S.~L.~D.}\ \bibnamefont {ten Haaf}}, \bibinfo {author} {\bibfnamefont {C.-X.}\ \bibnamefont {Liu}}, \bibinfo {author} {\bibfnamefont {F.}~\bibnamefont {Hassler}},\ and\ \bibinfo {author} {\bibfnamefont {M.}~\bibnamefont {Wimmer}},\ }\bibfield  {title} {\bibinfo {title} {{Interaction-induced strong zero modes in short quantum dot chains with time-reversal symmetry}},\ }\href {https://doi.org/10.21468/SciPostPhys.18.6.206} {\bibfield  {journal} {\bibinfo  {journal} {SciPost Phys.}\ }\textbf {\bibinfo {volume} {18}},\ \bibinfo {pages} {206} (\bibinfo {year} {2025})}\BibitemShut {NoStop}%
\bibitem [{\citenamefont {Fendley}(2012)}]{Fendley2012}%
  \BibitemOpen
  \bibfield  {author} {\bibinfo {author} {\bibfnamefont {P.}~\bibnamefont {Fendley}},\ }\bibfield  {title} {\bibinfo {title} {Parafermionic edge zero modes in {$Z_n$}-invariant spin chains},\ }\href {https://doi.org/10.1088/1742-5468/2012/11/P11020} {\bibfield  {journal} {\bibinfo  {journal} {Journal of Statistical Mechanics: Theory and Experiment}\ }\textbf {\bibinfo {volume} {2012}},\ \bibinfo {pages} {P11020} (\bibinfo {year} {2012})}\BibitemShut {NoStop}%
\bibitem [{\citenamefont {Alicea}\ and\ \citenamefont {Fendley}(2016)}]{Alicea2016}%
  \BibitemOpen
  \bibfield  {author} {\bibinfo {author} {\bibfnamefont {J.}~\bibnamefont {Alicea}}\ and\ \bibinfo {author} {\bibfnamefont {P.}~\bibnamefont {Fendley}},\ }\bibfield  {title} {\bibinfo {title} {Topological phases with parafermions: Theory and blueprints},\ }\href {https://doi.org/https://doi.org/10.1146/annurev-conmatphys-031115-011336} {\bibfield  {journal} {\bibinfo  {journal} {Annual Review of Condensed Matter Physics}\ }\textbf {\bibinfo {volume} {7}},\ \bibinfo {pages} {119} (\bibinfo {year} {2016})}\BibitemShut {NoStop}%
\bibitem [{\citenamefont {Wang}\ \emph {et~al.}(2022)\citenamefont {Wang}, \citenamefont {Dvir}, \citenamefont {Mazur}, \citenamefont {Liu}, \citenamefont {van Loo}, \citenamefont {ten Haaf}, \citenamefont {Bordin}, \citenamefont {Gazibegovic}, \citenamefont {Badawy}, \citenamefont {Bakkers}, \citenamefont {Wimmer},\ and\ \citenamefont {Kouwenhoven}}]{WangG2022}%
  \BibitemOpen
  \bibfield  {author} {\bibinfo {author} {\bibfnamefont {G.}~\bibnamefont {Wang}}, \bibinfo {author} {\bibfnamefont {T.}~\bibnamefont {Dvir}}, \bibinfo {author} {\bibfnamefont {G.~P.}\ \bibnamefont {Mazur}}, \bibinfo {author} {\bibfnamefont {C.-X.}\ \bibnamefont {Liu}}, \bibinfo {author} {\bibfnamefont {N.}~\bibnamefont {van Loo}}, \bibinfo {author} {\bibfnamefont {S.~L.~D.}\ \bibnamefont {ten Haaf}}, \bibinfo {author} {\bibfnamefont {A.}~\bibnamefont {Bordin}}, \bibinfo {author} {\bibfnamefont {S.}~\bibnamefont {Gazibegovic}}, \bibinfo {author} {\bibfnamefont {G.}~\bibnamefont {Badawy}}, \bibinfo {author} {\bibfnamefont {E.~P. A.~M.}\ \bibnamefont {Bakkers}}, \bibinfo {author} {\bibfnamefont {M.}~\bibnamefont {Wimmer}},\ and\ \bibinfo {author} {\bibfnamefont {L.~P.}\ \bibnamefont {Kouwenhoven}},\ }\bibfield  {title} {\bibinfo {title} {Singlet and triplet {Cooper} pair splitting in hybrid superconducting nanowires},\ }\href {https://doi.org/10.1038/s41586-022-05352-2} {\bibfield  {journal} {\bibinfo
  {journal} {Nature}\ }\textbf {\bibinfo {volume} {612}},\ \bibinfo {pages} {448} (\bibinfo {year} {2022})}\BibitemShut {NoStop}%
\bibitem [{\citenamefont {Bordin}\ \emph {et~al.}(2024)\citenamefont {Bordin}, \citenamefont {Li}, \citenamefont {van Driel}, \citenamefont {Wolff}, \citenamefont {Wang}, \citenamefont {ten Haaf}, \citenamefont {Wang}, \citenamefont {van Loo}, \citenamefont {Kouwenhoven},\ and\ \citenamefont {Dvir}}]{Bordin2024threesiteCAR}%
  \BibitemOpen
  \bibfield  {author} {\bibinfo {author} {\bibfnamefont {A.}~\bibnamefont {Bordin}}, \bibinfo {author} {\bibfnamefont {X.}~\bibnamefont {Li}}, \bibinfo {author} {\bibfnamefont {D.}~\bibnamefont {van Driel}}, \bibinfo {author} {\bibfnamefont {J.~C.}\ \bibnamefont {Wolff}}, \bibinfo {author} {\bibfnamefont {Q.}~\bibnamefont {Wang}}, \bibinfo {author} {\bibfnamefont {S.~L.~D.}\ \bibnamefont {ten Haaf}}, \bibinfo {author} {\bibfnamefont {G.}~\bibnamefont {Wang}}, \bibinfo {author} {\bibfnamefont {N.}~\bibnamefont {van Loo}}, \bibinfo {author} {\bibfnamefont {L.~P.}\ \bibnamefont {Kouwenhoven}},\ and\ \bibinfo {author} {\bibfnamefont {T.}~\bibnamefont {Dvir}},\ }\bibfield  {title} {\bibinfo {title} {Crossed andreev reflection and elastic cotunneling in three quantum dots coupled by superconductors},\ }\href {https://doi.org/10.1103/PhysRevLett.132.056602} {\bibfield  {journal} {\bibinfo  {journal} {Phys. Rev. Lett.}\ }\textbf {\bibinfo {volume} {132}},\ \bibinfo {pages} {056602} (\bibinfo {year}
  {2024})}\BibitemShut {NoStop}%
\bibitem [{\citenamefont {Bordin}\ \emph {et~al.}(2023)\citenamefont {Bordin}, \citenamefont {Wang}, \citenamefont {Liu}, \citenamefont {ten Haaf}, \citenamefont {van Loo}, \citenamefont {Mazur}, \citenamefont {Xu}, \citenamefont {van Driel}, \citenamefont {Zatelli}, \citenamefont {Gazibegovic}, \citenamefont {Badawy}, \citenamefont {Bakkers}, \citenamefont {Wimmer}, \citenamefont {Kouwenhoven},\ and\ \citenamefont {Dvir}}]{Bordin2023}%
  \BibitemOpen
  \bibfield  {author} {\bibinfo {author} {\bibfnamefont {A.}~\bibnamefont {Bordin}}, \bibinfo {author} {\bibfnamefont {G.}~\bibnamefont {Wang}}, \bibinfo {author} {\bibfnamefont {C.-X.}\ \bibnamefont {Liu}}, \bibinfo {author} {\bibfnamefont {S.~L.~D.}\ \bibnamefont {ten Haaf}}, \bibinfo {author} {\bibfnamefont {N.}~\bibnamefont {van Loo}}, \bibinfo {author} {\bibfnamefont {G.~P.}\ \bibnamefont {Mazur}}, \bibinfo {author} {\bibfnamefont {D.}~\bibnamefont {Xu}}, \bibinfo {author} {\bibfnamefont {D.}~\bibnamefont {van Driel}}, \bibinfo {author} {\bibfnamefont {F.}~\bibnamefont {Zatelli}}, \bibinfo {author} {\bibfnamefont {S.}~\bibnamefont {Gazibegovic}}, \bibinfo {author} {\bibfnamefont {G.}~\bibnamefont {Badawy}}, \bibinfo {author} {\bibfnamefont {E.~P. A.~M.}\ \bibnamefont {Bakkers}}, \bibinfo {author} {\bibfnamefont {M.}~\bibnamefont {Wimmer}}, \bibinfo {author} {\bibfnamefont {L.~P.}\ \bibnamefont {Kouwenhoven}},\ and\ \bibinfo {author} {\bibfnamefont {T.}~\bibnamefont {Dvir}},\ }\bibfield  {title} {\bibinfo
  {title} {Tunable crossed andreev reflection and elastic cotunneling in hybrid nanowires},\ }\href {https://doi.org/10.1103/PhysRevX.13.031031} {\bibfield  {journal} {\bibinfo  {journal} {Phys. Rev. X}\ }\textbf {\bibinfo {volume} {13}},\ \bibinfo {pages} {031031} (\bibinfo {year} {2023})}\BibitemShut {NoStop}%
\bibitem [{\citenamefont {Wang}\ \emph {et~al.}(2023)\citenamefont {Wang}, \citenamefont {ten Haaf}, \citenamefont {Kulesh}, \citenamefont {Xiao}, \citenamefont {Thomas}, \citenamefont {Manfra},\ and\ \citenamefont {Goswami}}]{WangQ2023}%
  \BibitemOpen
  \bibfield  {author} {\bibinfo {author} {\bibfnamefont {Q.}~\bibnamefont {Wang}}, \bibinfo {author} {\bibfnamefont {S.~L.~D.}\ \bibnamefont {ten Haaf}}, \bibinfo {author} {\bibfnamefont {I.}~\bibnamefont {Kulesh}}, \bibinfo {author} {\bibfnamefont {D.}~\bibnamefont {Xiao}}, \bibinfo {author} {\bibfnamefont {C.}~\bibnamefont {Thomas}}, \bibinfo {author} {\bibfnamefont {M.~J.}\ \bibnamefont {Manfra}},\ and\ \bibinfo {author} {\bibfnamefont {S.}~\bibnamefont {Goswami}},\ }\bibfield  {title} {\bibinfo {title} {Triplet correlations in {Cooper} pair splitters realized in a two-dimensional electron gas},\ }\href {https://doi.org/10.1038/s41467-023-40551-z} {\bibfield  {journal} {\bibinfo  {journal} {Nature Communications}\ }\textbf {\bibinfo {volume} {14}},\ \bibinfo {pages} {4876} (\bibinfo {year} {2023})}\BibitemShut {NoStop}%
\bibitem [{\citenamefont {O'Brien}\ \emph {et~al.}(2015)\citenamefont {O'Brien}, \citenamefont {Wright},\ and\ \citenamefont {Veldhorst}}]{OBrien2015}%
  \BibitemOpen
  \bibfield  {author} {\bibinfo {author} {\bibfnamefont {T.~E.}\ \bibnamefont {O'Brien}}, \bibinfo {author} {\bibfnamefont {A.~R.}\ \bibnamefont {Wright}},\ and\ \bibinfo {author} {\bibfnamefont {M.}~\bibnamefont {Veldhorst}},\ }\bibfield  {title} {\bibinfo {title} {Many-particle majorana bound states: Derivation and signatures in superconducting double quantum dots},\ }\href {https://doi.org/https://doi.org/10.1002/pssb.201552019} {\bibfield  {journal} {\bibinfo  {journal} {physica status solidi (b)}\ }\textbf {\bibinfo {volume} {252}},\ \bibinfo {pages} {1731} (\bibinfo {year} {2015})}\BibitemShut {NoStop}%
\bibitem [{\citenamefont {Wright}\ and\ \citenamefont {Veldhorst}(2013)}]{Wright2013}%
  \BibitemOpen
  \bibfield  {author} {\bibinfo {author} {\bibfnamefont {A.~R.}\ \bibnamefont {Wright}}\ and\ \bibinfo {author} {\bibfnamefont {M.}~\bibnamefont {Veldhorst}},\ }\bibfield  {title} {\bibinfo {title} {Localized many-particle majorana modes with vanishing time-reversal symmetry breaking in double quantum dots},\ }\href {https://doi.org/10.1103/PhysRevLett.111.096801} {\bibfield  {journal} {\bibinfo  {journal} {Phys. Rev. Lett.}\ }\textbf {\bibinfo {volume} {111}},\ \bibinfo {pages} {096801} (\bibinfo {year} {2013})}\BibitemShut {NoStop}%
\bibitem [{\citenamefont {Reilly}\ \emph {et~al.}(2007)\citenamefont {Reilly}, \citenamefont {Marcus}, \citenamefont {Hanson},\ and\ \citenamefont {Gossard}}]{reilly2007fast}%
  \BibitemOpen
  \bibfield  {author} {\bibinfo {author} {\bibfnamefont {D.}~\bibnamefont {Reilly}}, \bibinfo {author} {\bibfnamefont {C.}~\bibnamefont {Marcus}}, \bibinfo {author} {\bibfnamefont {M.}~\bibnamefont {Hanson}},\ and\ \bibinfo {author} {\bibfnamefont {A.}~\bibnamefont {Gossard}},\ }\bibfield  {title} {\bibinfo {title} {Fast single-charge sensing with a rf quantum point contact},\ }\bibfield  {journal} {\bibinfo  {journal} {Applied Physics Letters}\ }\textbf {\bibinfo {volume} {91}},\ \href {https://doi.org/10.1063/1.2794995} {10.1063/1.2794995} (\bibinfo {year} {2007})\BibitemShut {NoStop}%
\bibitem [{\citenamefont {Jung}\ \emph {et~al.}(2012)\citenamefont {Jung}, \citenamefont {Schroer}, \citenamefont {Petersson},\ and\ \citenamefont {Petta}}]{jung2012radio}%
  \BibitemOpen
  \bibfield  {author} {\bibinfo {author} {\bibfnamefont {M.}~\bibnamefont {Jung}}, \bibinfo {author} {\bibfnamefont {M.}~\bibnamefont {Schroer}}, \bibinfo {author} {\bibfnamefont {K.}~\bibnamefont {Petersson}},\ and\ \bibinfo {author} {\bibfnamefont {J.~R.}\ \bibnamefont {Petta}},\ }\bibfield  {title} {\bibinfo {title} {Radio frequency charge sensing in {InAs} nanowire double quantum dots},\ }\bibfield  {journal} {\bibinfo  {journal} {Applied Physics Letters}\ }\textbf {\bibinfo {volume} {100}},\ \href {https://doi.org/10.1063/1.4729469} {10.1063/1.4729469} (\bibinfo {year} {2012})\BibitemShut {NoStop}%
\bibitem [{\citenamefont {Razmadze}\ \emph {et~al.}(2019)\citenamefont {Razmadze}, \citenamefont {Sabonis}, \citenamefont {Malinowski}, \citenamefont {M\'enard}, \citenamefont {Pauka}, \citenamefont {Nguyen}, \citenamefont {van Zanten}, \citenamefont {O\ensuremath{'}Farrell}, \citenamefont {Suter}, \citenamefont {Krogstrup}, \citenamefont {Kuemmeth},\ and\ \citenamefont {Marcus}}]{razmadze2019radio}%
  \BibitemOpen
  \bibfield  {author} {\bibinfo {author} {\bibfnamefont {D.}~\bibnamefont {Razmadze}}, \bibinfo {author} {\bibfnamefont {D.}~\bibnamefont {Sabonis}}, \bibinfo {author} {\bibfnamefont {F.~K.}\ \bibnamefont {Malinowski}}, \bibinfo {author} {\bibfnamefont {G.~C.}\ \bibnamefont {M\'enard}}, \bibinfo {author} {\bibfnamefont {S.}~\bibnamefont {Pauka}}, \bibinfo {author} {\bibfnamefont {H.}~\bibnamefont {Nguyen}}, \bibinfo {author} {\bibfnamefont {D.~M.}\ \bibnamefont {van Zanten}}, \bibinfo {author} {\bibfnamefont {E.~C.}\ \bibnamefont {O\ensuremath{'}Farrell}}, \bibinfo {author} {\bibfnamefont {J.}~\bibnamefont {Suter}}, \bibinfo {author} {\bibfnamefont {P.}~\bibnamefont {Krogstrup}}, \bibinfo {author} {\bibfnamefont {F.}~\bibnamefont {Kuemmeth}},\ and\ \bibinfo {author} {\bibfnamefont {C.~M.}\ \bibnamefont {Marcus}},\ }\bibfield  {title} {\bibinfo {title} {Radio-frequency methods for {M}ajorana-based quantum devices: Fast charge sensing and phase-diagram mapping},\ }\href@noop {} {\bibfield  {journal} {\bibinfo
  {journal} {Phys. Rev. Appl.}\ }\textbf {\bibinfo {volume} {11}},\ \bibinfo {pages} {064011} (\bibinfo {year} {2019})}\BibitemShut {NoStop}%
\bibitem [{\citenamefont {Kulesh}\ \emph {et~al.}(2025)\citenamefont {Kulesh}, \citenamefont {ten Haaf}, \citenamefont {Wang}, \citenamefont {Sietses}, \citenamefont {Zhang}, \citenamefont {Roelofs}, \citenamefont {Prosko}, \citenamefont {Xiao}, \citenamefont {Thomas}, \citenamefont {Manfra},\ and\ \citenamefont {Goswami}}]{kulesh2025}%
  \BibitemOpen
  \bibfield  {author} {\bibinfo {author} {\bibfnamefont {I.}~\bibnamefont {Kulesh}}, \bibinfo {author} {\bibfnamefont {S.~L.~D.}\ \bibnamefont {ten Haaf}}, \bibinfo {author} {\bibfnamefont {Q.}~\bibnamefont {Wang}}, \bibinfo {author} {\bibfnamefont {V.~P.~M.}\ \bibnamefont {Sietses}}, \bibinfo {author} {\bibfnamefont {Y.}~\bibnamefont {Zhang}}, \bibinfo {author} {\bibfnamefont {S.~R.}\ \bibnamefont {Roelofs}}, \bibinfo {author} {\bibfnamefont {C.~G.}\ \bibnamefont {Prosko}}, \bibinfo {author} {\bibfnamefont {D.}~\bibnamefont {Xiao}}, \bibinfo {author} {\bibfnamefont {C.}~\bibnamefont {Thomas}}, \bibinfo {author} {\bibfnamefont {M.~J.}\ \bibnamefont {Manfra}},\ and\ \bibinfo {author} {\bibfnamefont {S.}~\bibnamefont {Goswami}},\ }\bibfield  {title} {\bibinfo {title} {Flux-controlled two-site {Kitaev} chain},\ }\href {https://doi.org/10.1103/r9pv-2prs} {\bibfield  {journal} {\bibinfo  {journal} {Phys. Rev. Lett.}\ }\textbf {\bibinfo {volume} {135}},\ \bibinfo {pages} {056301} (\bibinfo {year}
  {2025})}\BibitemShut {NoStop}%
\bibitem [{\citenamefont {Scherübl}\ \emph {et~al.}(2019)\citenamefont {Scherübl}, \citenamefont {Pályi},\ and\ \citenamefont {Csonka}}]{Scherubl2019}%
  \BibitemOpen
  \bibfield  {author} {\bibinfo {author} {\bibfnamefont {Z.}~\bibnamefont {Scherübl}}, \bibinfo {author} {\bibfnamefont {A.}~\bibnamefont {Pályi}},\ and\ \bibinfo {author} {\bibfnamefont {S.}~\bibnamefont {Csonka}},\ }\bibfield  {title} {\bibinfo {title} {Transport signatures of an {Andreev} molecule in a quantum dot–superconductor–quantum dot setup},\ }\href {https://doi.org/10.3762/bjnano.10.36} {\bibfield  {journal} {\bibinfo  {journal} {Beilstein Journal of Nanotechnology}\ }\textbf {\bibinfo {volume} {10}},\ \bibinfo {pages} {363} (\bibinfo {year} {2019})}\BibitemShut {NoStop}%
\bibitem [{\citenamefont {Dourado}\ \emph {et~al.}(2025)\citenamefont {Dourado}, \citenamefont {Leijnse},\ and\ \citenamefont {Souto}}]{Dourado2025}%
  \BibitemOpen
  \bibfield  {author} {\bibinfo {author} {\bibfnamefont {R.~A.}\ \bibnamefont {Dourado}}, \bibinfo {author} {\bibfnamefont {M.}~\bibnamefont {Leijnse}},\ and\ \bibinfo {author} {\bibfnamefont {R.~S.}\ \bibnamefont {Souto}},\ }\bibfield  {title} {\bibinfo {title} {Majorana sweet spots in three-site {Kitaev} chains},\ }\href {https://doi.org/10.1103/PhysRevB.111.235409} {\bibfield  {journal} {\bibinfo  {journal} {Phys. Rev. B}\ }\textbf {\bibinfo {volume} {111}},\ \bibinfo {pages} {235409} (\bibinfo {year} {2025})}\BibitemShut {NoStop}%
\bibitem [{\citenamefont {Liu}\ \emph {et~al.}(2025)\citenamefont {Liu}, \citenamefont {Miles}, \citenamefont {Bordin}, \citenamefont {ten Haaf}, \citenamefont {Mazur}, \citenamefont {Bozkurt},\ and\ \citenamefont {Wimmer}}]{cxliu2025}%
  \BibitemOpen
  \bibfield  {author} {\bibinfo {author} {\bibfnamefont {C.-X.}\ \bibnamefont {Liu}}, \bibinfo {author} {\bibfnamefont {S.}~\bibnamefont {Miles}}, \bibinfo {author} {\bibfnamefont {A.}~\bibnamefont {Bordin}}, \bibinfo {author} {\bibfnamefont {S.~L.~D.}\ \bibnamefont {ten Haaf}}, \bibinfo {author} {\bibfnamefont {G.~P.}\ \bibnamefont {Mazur}}, \bibinfo {author} {\bibfnamefont {A.~M.}\ \bibnamefont {Bozkurt}},\ and\ \bibinfo {author} {\bibfnamefont {M.}~\bibnamefont {Wimmer}},\ }\bibfield  {title} {\bibinfo {title} {Scaling up a sign-ordered kitaev chain without magnetic flux control},\ }\href {https://doi.org/10.1103/PhysRevResearch.7.L012045} {\bibfield  {journal} {\bibinfo  {journal} {Phys. Rev. Res.}\ }\textbf {\bibinfo {volume} {7}},\ \bibinfo {pages} {L012045} (\bibinfo {year} {2025})}\BibitemShut {NoStop}%
\bibitem [{\citenamefont {Jermyn}\ \emph {et~al.}(2014)\citenamefont {Jermyn}, \citenamefont {Mong}, \citenamefont {Alicea},\ and\ \citenamefont {Fendley}}]{jermyn2014}%
  \BibitemOpen
  \bibfield  {author} {\bibinfo {author} {\bibfnamefont {A.~S.}\ \bibnamefont {Jermyn}}, \bibinfo {author} {\bibfnamefont {R.~S.~K.}\ \bibnamefont {Mong}}, \bibinfo {author} {\bibfnamefont {J.}~\bibnamefont {Alicea}},\ and\ \bibinfo {author} {\bibfnamefont {P.}~\bibnamefont {Fendley}},\ }\bibfield  {title} {\bibinfo {title} {Stability of zero modes in parafermion chains},\ }\href {https://doi.org/10.1103/PhysRevB.90.165106} {\bibfield  {journal} {\bibinfo  {journal} {Phys. Rev. B}\ }\textbf {\bibinfo {volume} {90}},\ \bibinfo {pages} {165106} (\bibinfo {year} {2014})}\BibitemShut {NoStop}%
\bibitem [{\citenamefont {Prada}\ \emph {et~al.}(2017)\citenamefont {Prada}, \citenamefont {Aguado},\ and\ \citenamefont {San-Jose}}]{prada2017}%
  \BibitemOpen
  \bibfield  {author} {\bibinfo {author} {\bibfnamefont {E.}~\bibnamefont {Prada}}, \bibinfo {author} {\bibfnamefont {R.}~\bibnamefont {Aguado}},\ and\ \bibinfo {author} {\bibfnamefont {P.}~\bibnamefont {San-Jose}},\ }\bibfield  {title} {\bibinfo {title} {Measuring {Majorana} nonlocality and spin structure with a quantum dot},\ }\href {https://doi.org/10.1103/PhysRevB.96.085418} {\bibfield  {journal} {\bibinfo  {journal} {Phys. Rev. B}\ }\textbf {\bibinfo {volume} {96}},\ \bibinfo {pages} {085418} (\bibinfo {year} {2017})}\BibitemShut {NoStop}%
\bibitem [{\citenamefont {Clarke}(2017)}]{Clarke_2017}%
  \BibitemOpen
  \bibfield  {author} {\bibinfo {author} {\bibfnamefont {D.~J.}\ \bibnamefont {Clarke}},\ }\bibfield  {title} {\bibinfo {title} {Experimentally accessible topological quality factor for wires with zero energy modes},\ }\bibfield  {journal} {\bibinfo  {journal} {Physical Review B}\ }\textbf {\bibinfo {volume} {96}},\ \href {https://doi.org/10.1103/physrevb.96.201109} {10.1103/physrevb.96.201109} (\bibinfo {year} {2017})\BibitemShut {NoStop}%
\bibitem [{\citenamefont {Bordin}\ \emph {et~al.}(2025{\natexlab{b}})\citenamefont {Bordin}, \citenamefont {Evertsz'}, \citenamefont {Roovers}, \citenamefont {Luna}, \citenamefont {Huisman}, \citenamefont {Zatelli}, \citenamefont {Mazur}, \citenamefont {ten Haaf}, \citenamefont {Badawy}, \citenamefont {Bakkers}, \citenamefont {Liu}, \citenamefont {Souto}, \citenamefont {van Loo},\ and\ \citenamefont {Kouwenhoven}}]{bordin2025pradaclarke}%
  \BibitemOpen
  \bibfield  {author} {\bibinfo {author} {\bibfnamefont {A.}~\bibnamefont {Bordin}}, \bibinfo {author} {\bibfnamefont {F.~J.~B.}\ \bibnamefont {Evertsz'}}, \bibinfo {author} {\bibfnamefont {B.}~\bibnamefont {Roovers}}, \bibinfo {author} {\bibfnamefont {J.~D.~T.}\ \bibnamefont {Luna}}, \bibinfo {author} {\bibfnamefont {W.~D.}\ \bibnamefont {Huisman}}, \bibinfo {author} {\bibfnamefont {F.}~\bibnamefont {Zatelli}}, \bibinfo {author} {\bibfnamefont {G.~P.}\ \bibnamefont {Mazur}}, \bibinfo {author} {\bibfnamefont {S.~L.~D.}\ \bibnamefont {ten Haaf}}, \bibinfo {author} {\bibfnamefont {G.}~\bibnamefont {Badawy}}, \bibinfo {author} {\bibfnamefont {E.~P. A.~M.}\ \bibnamefont {Bakkers}}, \bibinfo {author} {\bibfnamefont {C.-X.}\ \bibnamefont {Liu}}, \bibinfo {author} {\bibfnamefont {R.~S.}\ \bibnamefont {Souto}}, \bibinfo {author} {\bibfnamefont {N.}~\bibnamefont {van Loo}},\ and\ \bibinfo {author} {\bibfnamefont {L.~P.}\ \bibnamefont {Kouwenhoven}},\ }\href {https://arxiv.org/abs/2504.13702} {\bibinfo {title} {Probing
  {Majorana} localization of a phase-controlled three-site {Kitaev} chain with an additional quantum dot}} (\bibinfo {year} {2025}{\natexlab{b}}),\ \Eprint {https://arxiv.org/abs/2504.13702} {arXiv:2504.13702 [cond-mat.mes-hall]} \BibitemShut {NoStop}%
\bibitem [{\citenamefont {Howes}\ \emph {et~al.}(1983)\citenamefont {Howes}, \citenamefont {Kadanoff},\ and\ \citenamefont {{Den Nijs}}}]{denNijs1983}%
  \BibitemOpen
  \bibfield  {author} {\bibinfo {author} {\bibfnamefont {S.}~\bibnamefont {Howes}}, \bibinfo {author} {\bibfnamefont {L.~P.}\ \bibnamefont {Kadanoff}},\ and\ \bibinfo {author} {\bibfnamefont {M.}~\bibnamefont {{Den Nijs}}},\ }\bibfield  {title} {\bibinfo {title} {Quantum model for commensurate-incommensurate transitions},\ }\href {https://doi.org/https://doi.org/10.1016/0550-3213(83)90212-2} {\bibfield  {journal} {\bibinfo  {journal} {Nuclear Physics B}\ }\textbf {\bibinfo {volume} {215}},\ \bibinfo {pages} {169} (\bibinfo {year} {1983})}\BibitemShut {NoStop}%
\bibitem [{\citenamefont {Ostlund}(1981)}]{Ostlund1981}%
  \BibitemOpen
  \bibfield  {author} {\bibinfo {author} {\bibfnamefont {S.}~\bibnamefont {Ostlund}},\ }\bibfield  {title} {\bibinfo {title} {Incommensurate and commensurate phases in asymmetric clock models},\ }\href {https://doi.org/10.1103/PhysRevB.24.398} {\bibfield  {journal} {\bibinfo  {journal} {Phys. Rev. B}\ }\textbf {\bibinfo {volume} {24}},\ \bibinfo {pages} {398} (\bibinfo {year} {1981})}\BibitemShut {NoStop}%
\bibitem [{\citenamefont {Teixeira}\ and\ \citenamefont {Dias~da Silva}(2022)}]{teixeria2022}%
  \BibitemOpen
  \bibfield  {author} {\bibinfo {author} {\bibfnamefont {R.~L. R.~C.}\ \bibnamefont {Teixeira}}\ and\ \bibinfo {author} {\bibfnamefont {L.~G. G.~V.}\ \bibnamefont {Dias~da Silva}},\ }\bibfield  {title} {\bibinfo {title} {Edge ${\mathbb{z}}_{3}$ parafermions in fermionic lattices},\ }\href {https://doi.org/10.1103/PhysRevB.105.195121} {\bibfield  {journal} {\bibinfo  {journal} {Phys. Rev. B}\ }\textbf {\bibinfo {volume} {105}},\ \bibinfo {pages} {195121} (\bibinfo {year} {2022})}\BibitemShut {NoStop}%
\bibitem [{\citenamefont {Day}\ \emph {et~al.}(2025)\citenamefont {Day}, \citenamefont {Miles}, \citenamefont {Kerstens}, \citenamefont {Varjas},\ and\ \citenamefont {Akhmerov}}]{araya2025}%
  \BibitemOpen
  \bibfield  {author} {\bibinfo {author} {\bibfnamefont {I.~A.}\ \bibnamefont {Day}}, \bibinfo {author} {\bibfnamefont {S.}~\bibnamefont {Miles}}, \bibinfo {author} {\bibfnamefont {H.~K.}\ \bibnamefont {Kerstens}}, \bibinfo {author} {\bibfnamefont {D.}~\bibnamefont {Varjas}},\ and\ \bibinfo {author} {\bibfnamefont {A.~R.}\ \bibnamefont {Akhmerov}},\ }\bibfield  {title} {\bibinfo {title} {{Codebase release 2.1 for Pymablock}},\ }\href {https://doi.org/10.21468/SciPostPhysCodeb.50-r2.1} {\bibfield  {journal} {\bibinfo  {journal} {SciPost Phys. Codebases}\ ,\ \bibinfo {pages} {50}} (\bibinfo {year} {2025})}\BibitemShut {NoStop}%
\bibitem [{\citenamefont {Gao}\ \emph {et~al.}(2016)\citenamefont {Gao}, \citenamefont {He},\ and\ \citenamefont {Liu}}]{Gao2016}%
  \BibitemOpen
  \bibfield  {author} {\bibinfo {author} {\bibfnamefont {P.}~\bibnamefont {Gao}}, \bibinfo {author} {\bibfnamefont {Y.-P.}\ \bibnamefont {He}},\ and\ \bibinfo {author} {\bibfnamefont {X.-J.}\ \bibnamefont {Liu}},\ }\bibfield  {title} {\bibinfo {title} {Symmetry-protected non-{Abelian} braiding of {Majorana} {Kramers} pairs},\ }\href {https://doi.org/10.1103/PhysRevB.94.224509} {\bibfield  {journal} {\bibinfo  {journal} {Phys. Rev. B}\ }\textbf {\bibinfo {volume} {94}},\ \bibinfo {pages} {224509} (\bibinfo {year} {2016})}\BibitemShut {NoStop}%
\bibitem [{\citenamefont {Schrade}\ and\ \citenamefont {Fu}(2022)}]{Schrade2022}%
  \BibitemOpen
  \bibfield  {author} {\bibinfo {author} {\bibfnamefont {C.}~\bibnamefont {Schrade}}\ and\ \bibinfo {author} {\bibfnamefont {L.}~\bibnamefont {Fu}},\ }\bibfield  {title} {\bibinfo {title} {Quantum computing with {Majorana} {Kramers} pairs},\ }\href {https://doi.org/10.1103/PhysRevLett.129.227002} {\bibfield  {journal} {\bibinfo  {journal} {Phys. Rev. Lett.}\ }\textbf {\bibinfo {volume} {129}},\ \bibinfo {pages} {227002} (\bibinfo {year} {2022})}\BibitemShut {NoStop}%
\bibitem [{\citenamefont {Pan}\ \emph {et~al.}(2025)\citenamefont {Pan}, \citenamefont {Jia},\ and\ \citenamefont {Qiao}}]{Pan2025}%
  \BibitemOpen
  \bibfield  {author} {\bibinfo {author} {\bibfnamefont {H.}~\bibnamefont {Pan}}, \bibinfo {author} {\bibfnamefont {J.}~\bibnamefont {Jia}},\ and\ \bibinfo {author} {\bibfnamefont {Z.}~\bibnamefont {Qiao}},\ }\bibfield  {title} {\bibinfo {title} {Fusion rules for {Majorana} {Kramers} pairs in time-reversal invariant topological superconductors},\ }\href {https://doi.org/10.1103/PhysRevB.111.115414} {\bibfield  {journal} {\bibinfo  {journal} {Phys. Rev. B}\ }\textbf {\bibinfo {volume} {111}},\ \bibinfo {pages} {115414} (\bibinfo {year} {2025})}\BibitemShut {NoStop}%
\bibitem [{\citenamefont {Wong}\ and\ \citenamefont {Law}(2012)}]{Wong2012}%
  \BibitemOpen
  \bibfield  {author} {\bibinfo {author} {\bibfnamefont {C.~L.~M.}\ \bibnamefont {Wong}}\ and\ \bibinfo {author} {\bibfnamefont {K.~T.}\ \bibnamefont {Law}},\ }\bibfield  {title} {\bibinfo {title} {Majorana kramers doublets in ${d}_{{x}^{2}\ensuremath{-}{y}^{2}}$-wave superconductors with rashba spin-orbit coupling},\ }\href {https://doi.org/10.1103/PhysRevB.86.184516} {\bibfield  {journal} {\bibinfo  {journal} {Phys. Rev. B}\ }\textbf {\bibinfo {volume} {86}},\ \bibinfo {pages} {184516} (\bibinfo {year} {2012})}\BibitemShut {NoStop}%
\bibitem [{\citenamefont {Zhang}\ \emph {et~al.}(2013)\citenamefont {Zhang}, \citenamefont {Kane},\ and\ \citenamefont {Mele}}]{FZhang2013}%
  \BibitemOpen
  \bibfield  {author} {\bibinfo {author} {\bibfnamefont {F.}~\bibnamefont {Zhang}}, \bibinfo {author} {\bibfnamefont {C.~L.}\ \bibnamefont {Kane}},\ and\ \bibinfo {author} {\bibfnamefont {E.~J.}\ \bibnamefont {Mele}},\ }\bibfield  {title} {\bibinfo {title} {Time-reversal-invariant topological superconductivity and {Majorana} {Kramers} pairs},\ }\href {https://doi.org/10.1103/PhysRevLett.111.056402} {\bibfield  {journal} {\bibinfo  {journal} {Phys. Rev. Lett.}\ }\textbf {\bibinfo {volume} {111}},\ \bibinfo {pages} {056402} (\bibinfo {year} {2013})}\BibitemShut {NoStop}%
\bibitem [{\citenamefont {Keselman}\ \emph {et~al.}(2013)\citenamefont {Keselman}, \citenamefont {Fu}, \citenamefont {Stern},\ and\ \citenamefont {Berg}}]{Keselman2013}%
  \BibitemOpen
  \bibfield  {author} {\bibinfo {author} {\bibfnamefont {A.}~\bibnamefont {Keselman}}, \bibinfo {author} {\bibfnamefont {L.}~\bibnamefont {Fu}}, \bibinfo {author} {\bibfnamefont {A.}~\bibnamefont {Stern}},\ and\ \bibinfo {author} {\bibfnamefont {E.}~\bibnamefont {Berg}},\ }\bibfield  {title} {\bibinfo {title} {Inducing time-reversal-invariant topological superconductivity and fermion parity pumping in quantum wires},\ }\href {https://doi.org/10.1103/PhysRevLett.111.116402} {\bibfield  {journal} {\bibinfo  {journal} {Phys. Rev. Lett.}\ }\textbf {\bibinfo {volume} {111}},\ \bibinfo {pages} {116402} (\bibinfo {year} {2013})}\BibitemShut {NoStop}%
\bibitem [{\citenamefont {Klinovaja}\ and\ \citenamefont {Loss}(2014)}]{Klinovaja2014}%
  \BibitemOpen
  \bibfield  {author} {\bibinfo {author} {\bibfnamefont {J.}~\bibnamefont {Klinovaja}}\ and\ \bibinfo {author} {\bibfnamefont {D.}~\bibnamefont {Loss}},\ }\bibfield  {title} {\bibinfo {title} {Time-reversal invariant parafermions in interacting {Rashba} nanowires},\ }\href {https://doi.org/10.1103/PhysRevB.90.045118} {\bibfield  {journal} {\bibinfo  {journal} {Phys. Rev. B}\ }\textbf {\bibinfo {volume} {90}},\ \bibinfo {pages} {045118} (\bibinfo {year} {2014})}\BibitemShut {NoStop}%
\bibitem [{\citenamefont {Klinovaja}\ \emph {et~al.}(2014)\citenamefont {Klinovaja}, \citenamefont {Yacoby},\ and\ \citenamefont {Loss}}]{Klinovaja2014b}%
  \BibitemOpen
  \bibfield  {author} {\bibinfo {author} {\bibfnamefont {J.}~\bibnamefont {Klinovaja}}, \bibinfo {author} {\bibfnamefont {A.}~\bibnamefont {Yacoby}},\ and\ \bibinfo {author} {\bibfnamefont {D.}~\bibnamefont {Loss}},\ }\bibfield  {title} {\bibinfo {title} {Kramers pairs of majorana fermions and parafermions in fractional topological insulators},\ }\href {https://doi.org/10.1103/PhysRevB.90.155447} {\bibfield  {journal} {\bibinfo  {journal} {Phys. Rev. B}\ }\textbf {\bibinfo {volume} {90}},\ \bibinfo {pages} {155447} (\bibinfo {year} {2014})}\BibitemShut {NoStop}%
\bibitem [{\citenamefont {Haim}\ \emph {et~al.}(2014)\citenamefont {Haim}, \citenamefont {Keselman}, \citenamefont {Berg},\ and\ \citenamefont {Oreg}}]{Haim2014}%
  \BibitemOpen
  \bibfield  {author} {\bibinfo {author} {\bibfnamefont {A.}~\bibnamefont {Haim}}, \bibinfo {author} {\bibfnamefont {A.}~\bibnamefont {Keselman}}, \bibinfo {author} {\bibfnamefont {E.}~\bibnamefont {Berg}},\ and\ \bibinfo {author} {\bibfnamefont {Y.}~\bibnamefont {Oreg}},\ }\bibfield  {title} {\bibinfo {title} {Time-reversal-invariant topological superconductivity induced by repulsive interactions in quantum wires},\ }\href {https://doi.org/10.1103/PhysRevB.89.220504} {\bibfield  {journal} {\bibinfo  {journal} {Phys. Rev. B}\ }\textbf {\bibinfo {volume} {89}},\ \bibinfo {pages} {220504} (\bibinfo {year} {2014})}\BibitemShut {NoStop}%
\bibitem [{\citenamefont {Liu}\ \emph {et~al.}(2014)\citenamefont {Liu}, \citenamefont {Wong},\ and\ \citenamefont {Law}}]{XJLiu2014}%
  \BibitemOpen
  \bibfield  {author} {\bibinfo {author} {\bibfnamefont {X.-J.}\ \bibnamefont {Liu}}, \bibinfo {author} {\bibfnamefont {C.~L.~M.}\ \bibnamefont {Wong}},\ and\ \bibinfo {author} {\bibfnamefont {K.~T.}\ \bibnamefont {Law}},\ }\bibfield  {title} {\bibinfo {title} {Non-abelian {Majorana} doublets in time-reversal-invariant topological superconductors},\ }\href {https://doi.org/10.1103/PhysRevX.4.021018} {\bibfield  {journal} {\bibinfo  {journal} {Phys. Rev. X}\ }\textbf {\bibinfo {volume} {4}},\ \bibinfo {pages} {021018} (\bibinfo {year} {2014})}\BibitemShut {NoStop}%
\bibitem [{\citenamefont {Schrade}\ \emph {et~al.}(2015)\citenamefont {Schrade}, \citenamefont {Zyuzin}, \citenamefont {Klinovaja},\ and\ \citenamefont {Loss}}]{Schrade2015}%
  \BibitemOpen
  \bibfield  {author} {\bibinfo {author} {\bibfnamefont {C.}~\bibnamefont {Schrade}}, \bibinfo {author} {\bibfnamefont {A.~A.}\ \bibnamefont {Zyuzin}}, \bibinfo {author} {\bibfnamefont {J.}~\bibnamefont {Klinovaja}},\ and\ \bibinfo {author} {\bibfnamefont {D.}~\bibnamefont {Loss}},\ }\bibfield  {title} {\bibinfo {title} {Proximity-induced $\ensuremath{\pi}$ josephson junctions in topological insulators and kramers pairs of majorana fermions},\ }\href {https://doi.org/10.1103/PhysRevLett.115.237001} {\bibfield  {journal} {\bibinfo  {journal} {Phys. Rev. Lett.}\ }\textbf {\bibinfo {volume} {115}},\ \bibinfo {pages} {237001} (\bibinfo {year} {2015})}\BibitemShut {NoStop}%
\bibitem [{\citenamefont {Thakurathi}\ \emph {et~al.}(2018)\citenamefont {Thakurathi}, \citenamefont {Simon}, \citenamefont {Mandal}, \citenamefont {Klinovaja},\ and\ \citenamefont {Loss}}]{Thakurathi2018}%
  \BibitemOpen
  \bibfield  {author} {\bibinfo {author} {\bibfnamefont {M.}~\bibnamefont {Thakurathi}}, \bibinfo {author} {\bibfnamefont {P.}~\bibnamefont {Simon}}, \bibinfo {author} {\bibfnamefont {I.}~\bibnamefont {Mandal}}, \bibinfo {author} {\bibfnamefont {J.}~\bibnamefont {Klinovaja}},\ and\ \bibinfo {author} {\bibfnamefont {D.}~\bibnamefont {Loss}},\ }\bibfield  {title} {\bibinfo {title} {Majorana kramers pairs in rashba double nanowires with interactions and disorder},\ }\href {https://doi.org/10.1103/PhysRevB.97.045415} {\bibfield  {journal} {\bibinfo  {journal} {Phys. Rev. B}\ }\textbf {\bibinfo {volume} {97}},\ \bibinfo {pages} {045415} (\bibinfo {year} {2018})}\BibitemShut {NoStop}%
\bibitem [{\citenamefont {ten Haaf}\ and\ \citenamefont {Miles}(2024)}]{ZenodoDataRepo}%
  \BibitemOpen
  \bibfield  {author} {\bibinfo {author} {\bibfnamefont {S.~L.~D.}\ \bibnamefont {ten Haaf}}\ and\ \bibinfo {author} {\bibfnamefont {S.}~\bibnamefont {Miles}},\ }\href {https://zenodo.org/doi/10.5281/zenodo.17872114} {\bibinfo {title} {Data and {Code} for ``{Probing} ground-state degeneracies of a strongly interacting {Fermi-Hubbard} model with superconducting correlations"}} (\bibinfo {year} {2024}),\ \bibinfo {note} {https://zenodo.org/doi/10.5281/zenodo.17872114}\BibitemShut {NoStop}%
\bibitem [{\citenamefont {Moehle}\ \emph {et~al.}(2021)\citenamefont {Moehle}, \citenamefont {Ke}, \citenamefont {Wang}, \citenamefont {Thomas}, \citenamefont {Xiao}, \citenamefont {Karwal}, \citenamefont {Lodari}, \citenamefont {van~de Kerkhof}, \citenamefont {Termaat}, \citenamefont {Gardner}, \citenamefont {Scappucci}, \citenamefont {Manfra},\ and\ \citenamefont {Goswami}}]{Moehle2021}%
  \BibitemOpen
  \bibfield  {author} {\bibinfo {author} {\bibfnamefont {C.~M.}\ \bibnamefont {Moehle}}, \bibinfo {author} {\bibfnamefont {C.~T.}\ \bibnamefont {Ke}}, \bibinfo {author} {\bibfnamefont {Q.}~\bibnamefont {Wang}}, \bibinfo {author} {\bibfnamefont {C.}~\bibnamefont {Thomas}}, \bibinfo {author} {\bibfnamefont {D.}~\bibnamefont {Xiao}}, \bibinfo {author} {\bibfnamefont {S.}~\bibnamefont {Karwal}}, \bibinfo {author} {\bibfnamefont {M.}~\bibnamefont {Lodari}}, \bibinfo {author} {\bibfnamefont {V.}~\bibnamefont {van~de Kerkhof}}, \bibinfo {author} {\bibfnamefont {R.}~\bibnamefont {Termaat}}, \bibinfo {author} {\bibfnamefont {G.~C.}\ \bibnamefont {Gardner}}, \bibinfo {author} {\bibfnamefont {G.}~\bibnamefont {Scappucci}}, \bibinfo {author} {\bibfnamefont {M.~J.}\ \bibnamefont {Manfra}},\ and\ \bibinfo {author} {\bibfnamefont {S.}~\bibnamefont {Goswami}},\ }\bibfield  {title} {\bibinfo {title} {{InSbAs} two-dimensional electron gases as a platform for topological superconductivity},\ }\href
  {https://doi.org/10.1021/acs.nanolett.1c03520} {\bibfield  {journal} {\bibinfo  {journal} {Nano Letters}\ }\textbf {\bibinfo {volume} {21}},\ \bibinfo {pages} {9990} (\bibinfo {year} {2021})}\BibitemShut {NoStop}%
\bibitem [{\citenamefont {Kulesh}(2025)}]{KuleshIThesis}%
  \BibitemOpen
  \bibfield  {author} {\bibinfo {author} {\bibfnamefont {I.}~\bibnamefont {Kulesh}},\ }\emph {\bibinfo {title} {Superconducting proximity and confinement in a two-dimensional electron gas.}},\ \href {https://doi.org/10.4233/uuid:41f57e3e-4163-4fdc-997f-f4f4b1e8fde1} {Ph.D. thesis},\ \bibinfo  {school} {Delft University of technology} (\bibinfo {year} {2025})\BibitemShut {NoStop}%
\bibitem [{\citenamefont {Wang}(2025)}]{WangQThesis}%
  \BibitemOpen
  \bibfield  {author} {\bibinfo {author} {\bibfnamefont {Q.}~\bibnamefont {Wang}},\ }\emph {\bibinfo {title} {Topological superconductivity in {InSbAs} two-dimensional electron gases.}},\ \href@noop {} {Ph.D. thesis},\ \bibinfo  {school} {Delft University of technology} (\bibinfo {year} {2025}),\ \bibinfo {note} {in prepration}\BibitemShut {NoStop}%
\bibitem [{\citenamefont {Vigneau}\ \emph {et~al.}(2023)\citenamefont {Vigneau}, \citenamefont {Fedele}, \citenamefont {Chatterjee}, \citenamefont {Reilly}, \citenamefont {Kuemmeth}, \citenamefont {Gonzalez-Zalba}, \citenamefont {Laird},\ and\ \citenamefont {Ares}}]{Vigneau2023}%
  \BibitemOpen
  \bibfield  {author} {\bibinfo {author} {\bibfnamefont {F.}~\bibnamefont {Vigneau}}, \bibinfo {author} {\bibfnamefont {F.}~\bibnamefont {Fedele}}, \bibinfo {author} {\bibfnamefont {A.}~\bibnamefont {Chatterjee}}, \bibinfo {author} {\bibfnamefont {D.}~\bibnamefont {Reilly}}, \bibinfo {author} {\bibfnamefont {F.}~\bibnamefont {Kuemmeth}}, \bibinfo {author} {\bibfnamefont {M.~F.}\ \bibnamefont {Gonzalez-Zalba}}, \bibinfo {author} {\bibfnamefont {E.}~\bibnamefont {Laird}},\ and\ \bibinfo {author} {\bibfnamefont {N.}~\bibnamefont {Ares}},\ }\bibfield  {title} {\bibinfo {title} {{Probing quantum devices with radio-frequency reflectometry}},\ }\href {https://doi.org/10.1063/5.0088229} {\bibfield  {journal} {\bibinfo  {journal} {Applied Physics Reviews}\ }\textbf {\bibinfo {volume} {10}},\ \bibinfo {pages} {021305} (\bibinfo {year} {2023})}\BibitemShut {NoStop}%
\bibitem [{\citenamefont {Hornibrook}\ \emph {et~al.}(2014)\citenamefont {Hornibrook}, \citenamefont {Colless}, \citenamefont {Mahoney}, \citenamefont {Croot}, \citenamefont {Blanvillain}, \citenamefont {Lu}, \citenamefont {Gossard},\ and\ \citenamefont {Reilly}}]{Hornibrook2014}%
  \BibitemOpen
  \bibfield  {author} {\bibinfo {author} {\bibfnamefont {J.}~\bibnamefont {Hornibrook}}, \bibinfo {author} {\bibfnamefont {J.}~\bibnamefont {Colless}}, \bibinfo {author} {\bibfnamefont {A.}~\bibnamefont {Mahoney}}, \bibinfo {author} {\bibfnamefont {X.}~\bibnamefont {Croot}}, \bibinfo {author} {\bibfnamefont {S.}~\bibnamefont {Blanvillain}}, \bibinfo {author} {\bibfnamefont {H.}~\bibnamefont {Lu}}, \bibinfo {author} {\bibfnamefont {A.}~\bibnamefont {Gossard}},\ and\ \bibinfo {author} {\bibfnamefont {D.}~\bibnamefont {Reilly}},\ }\bibfield  {title} {\bibinfo {title} {Frequency multiplexing for readout of spin qubits},\ }\bibfield  {journal} {\bibinfo  {journal} {Applied Physics Letters}\ }\textbf {\bibinfo {volume} {104}},\ \href {https://doi.org/10.1063/1.4868107} {10.1063/1.4868107} (\bibinfo {year} {2014})\BibitemShut {NoStop}%
\end{thebibliography}%

\newpage
\onecolumngrid

\appendix

\newcommand{\beginsupplement}{%
	\setcounter{section}{0}
	\renewcommand{\thesection}{\arabic{section}}%
	\setcounter{table}{0}
	\renewcommand{\thetable}{S\arabic{table}}%
	\setcounter{figure}{0}
	\renewcommand{\thefigure}
    {S\arabic{figure}}%
	\renewcommand{\theHfigure}{S\arabic{figure}}%
	\setcounter{equation}{0}
	\renewcommand{\theequation}{S\arabic{equation}}%
    \renewcommand{\theHequation}{S\arabic{equation}}%
}
\setcounter{subsection}{1}
\beginsupplement
\section*{Methods: Experiment}
\subsection{Fabrication and yield}
The presented devices are fabricated on InSbAS with \SI{7}{\nano\meter} epitaxial aluminium, described in detail in~\cite{Moehle2021}. 
An in-depth account of the fabrication of these specific devices can be found in \cite{KuleshIThesis} and \cite{WangQThesis}. 
Starting with an InSbAs chip fully covered with aluminium, a Transene D wet etch is used to create the fine structures, followed by the deposition of ohmic Ti/Pd contacts.
Next, we deposit \SI{20}{\nano\meter} AlOx via \SI{40}{\celsius} atomic layer deposition (ALD). 
To confine a quasi one dimensional channel, large Ti/Pd depletion gates are evaporated. 
Device B consists of two depletion gates (one top and one bottom), whereas device A utilises three depletion gate (one top and two bottom).
The extra bottom depletion gate simplifies utilising the additional middle ohmic contact and gives independent control over forming the left and right sides of the channel. 
The channel widths were designed to be $\approx$\SI{200}{\nano\meter}.
Following a second ALD layer (\SI{20}{\nano\meter} AlOx), we deposit a layer of Ti/Pd finger gates used to control the electrochemical potential energies of the QDs and the hybrid regions.
Lastly, a third ALD layer (\SI{20}{\nano\meter} AlOx) is deposited, followed by the final Ti/Pd tunnelling gates used to define the quantum dots.
Superconducting LC-resonator circuits were fabricated on a separate chip with a silicon substrate, by etching NbTiN.
To apply DC voltages, bias tees are created by depositing \SI{20}{\nano\meter} Cr structures with resistances of $\approx$~\SI{5}{\kilo\ohm}.
Further details on the resonator circuits and techniques can be found in~\cite{KuleshIThesis}.

\subsection{DC transport and RF-reflectometry measurements}
To efficiently explore the large parameter space of device A, we employed radio frequency (RF) lead reflectometry~\cite{Vigneau2023} in addition to DC conductance measurements.
To do so, we connect each ohmic contact to an inductor chip via a bond wire.
The inductors are designed with inductances $L_{\mathrm{1, 2, 3}} = 0.2, 0.5, 1.5$~\SI{}{\micro \henry}, that together with a parasitic capacitance to ground via bond-wires result in resonators with frequencies of $f_{\mathrm{1, 2, 3}} = 723, 505, 248$~\SI{}{\mega\hertz}.
To utilise the fast integration times offered by the resonator circuit, arbitrary waveform generators (AWGs) are used to vary voltages on either the QD plunger gates or to apply a bias on the ohmic contacts using sawtooth pulses with frequencies on the order of 10-\SI{50}{\hertz}.
A complete overview of the circuit diagram, fridge wirings and filters can be found in~\cite{kulesh2025,KuleshIThesis}.
Using a cryogenic directional coupler, we obtain simultaneously the reflected signal of each resonator through multiplexing~\cite{Hornibrook2014}.
For each resonator, both the amplitude and phase response are recorded. 
We convert this to a single normalized value for each lead (denoted $\tilde{S}_1$, $\tilde{S}_2$ and $\tilde{S}_3$), which corresponds roughly linearly to the conductance through each lead~\cite{reilly2007fast, jung2012radio,razmadze2019radio}.
The data processing procedure is shown in \cref{fig:S2_RF_processing}.
\newline

Device B was not connected to a resonator chip and only measured using DC and AC transport techniques. The same techniques were applied to device A, with connections routed via \SI{5}{\kilo\ohm} Cr structures, to isolate the DC lines from the RF-circuit.
In both devices, the aluminium strips are kept electrically grounded.
Each lead is connected to a current meter, which is biased with a digital-to-analogue converter, connected such that both DC and AC voltages can be applied.
The offsets of the applied voltage-biases on each lead are corrected via independently measuring the Coulomb peaks in the QDs and looking at the change in sign of the current as a function of applied voltage.
The voltage outputs of the current meters are each recorded with both a digital multimeter and a lock-in amplifier.  
When applying a DC/AC voltage to one lead, all other leads are kept grounded.
The AC excitations are applied with frequencies around \SI{20}{\hertz} and an amplitude of \SI{5}{\micro\volt} RMS.
Full conductance matrices $G_{ij}=\frac{dI_{i}}{dV_{j}}$ are obtained by measuring the response of each \var{I}{i} to each \var{V}{i}.
Typically, voltage-divider effects arise when applying biases in a multi-terminal set-up.
In these measurements we focus on low tunneling regimes (G $\ll$ 2e$^2$/h), such that the device resistance is significant compared to the resistance of other connections to ground and hence we do not correct these multi-terminal effects.
When measuring in RF, conductance measurements are first obtained to confirm whether we were indeed in the low-tunneling regime.

For the measurements in device A, the field perpendicular to the superconducting loop (\var{B}{z}) is generated using a high-resolution current source, providing a \var{B}{z} resolution below \SI{0.1}{\micro\tesla}. 
A small (but significant) hysteresis on the order of \SI{5}{\micro\tesla} is observed when sweeping \var{B}{z} in opposite directions.
This is counteracted by setting \var{B}{z} first to \SI{-100}{\micro\tesla} and then sweeping this field back in the positive direction, such that consecutive experiments where \var{B}{z} is varied are consistent.

\section*{Methods: Theory}

\subsection{Analytic spectrum along triply degenerate line}\label{app:analytic_spectrum}
In the limit $U\rightarrow \infty$ the Hilbert space is restricted to states only containing at most single occupation per site.
For $\mu_{i,\sigma}=0$ this restricted low energy spectrum can be calculated analytically, which we will show in the remainder of this section.
With the doubly occupied states unavailable, the size of the Hilbert space reduces from 64 down to 27, 13 even and 14 odd states.
As outlined in the main text, the Hamiltonian also commutes with the fermion parity operator $[H,P_F]=0$.
Hence, we can separate the Hilbert space into an odd and even fermion parity sector and solve them independently of each other.

\subsubsection{Even parity sector}
In the even parity sector we find that the Hamiltonian can be split into three blocks.
Beginning with the spin mixing block we find

\begin{align}
    H_{even}^{(1)}=\begin{pmatrix}
    0 & \Delta_{1} e^{- i \varphi_{1}} & 0 & \Delta_{0} & - \Delta_{1} e^{- i \varphi_{1}} & 0 & - \Delta_{0}\\\Delta_{1} e^{i \varphi_{1}} & 0 & t_{0} & 0 & 0 & 0 & 0\\0 & t_{0} & 0 & t_{1} & 0 & 0 & 0\\\Delta_{0} & 0 & t_{1} & 0 & 0 & 0 & 0\\- \Delta_{1} e^{i \varphi_{1}} & 0 & 0 & 0 & 0 & t_{0} & 0\\0 & 0 & 0 & 0 & t_{0} & 0 & t_{1}\\- \Delta_{0} & 0 & 0 & 0 & 0 & t_{1} & 0
    \end{pmatrix},
\end{align}
spanned by the states $\{|000\rangle, |0\mathord{\ua}\mathord{\da}\rangle,|\mathord{\ua}0\mathord{\da}\rangle,|\mathord{\ua}\mathord{\da}0\rangle,|0\mathord{\da}\mathord{\ua}\rangle,|\mathord{\da}0\mathord{\ua}\rangle,|\mathord{\da}\mathord{\ua}0\rangle\}$.
Calculation of the seventh order characteristic polynomial reveals:

\begin{align}
E_{even,1}^{(1)}=0.
\end{align}

Due to time-reversal symmetry, the remainder of the subblock spectrum is twofold degenerate.
We obtain:

\begin{align}
    E_{even,2}^{(1)}&=\sqrt{t_0^2+t_1^2},  \\
    E_{even,3}^{(1)}&=-\sqrt{t_0^2+t_1^2}  \\
    E_{even,4}^{(1)} &= \left(\Delta_0^2+\Delta_1^2+\frac{t_0^2+t_1^2}{2}+\left((\Delta_0^2+\Delta_1^2)^2+\frac{(t_0^2+t_1^2)^2}{4}+(\Delta_0^2-\Delta_1^2)(t_1^2-t_0^2)+4\Delta_0\Delta_1t_0t_1\cos(\varphi)\right)^{\frac{1}{2}}\right)^{\frac{1}{2}} \\
        E_{even,5}^{(1)} &= -\left(\Delta_0^2+\Delta_1^2+\frac{t_0^2+t_1^2}{2}+\left((\Delta_0^2+\Delta_1^2)^2+\frac{(t_0^2+t_1^2)^2}{4}+(\Delta_0^2-\Delta_1^2)(t_1^2-t_0^2)+4\Delta_0\Delta_1t_0t_1\cos(\varphi)\right)^{\frac{1}{2}}\right)^{\frac{1}{2}} \\
            E_{even,6}^{(1)} &= \left(\Delta_0^2+\Delta_1^2+\frac{t_0^2+t_1^2}{2}-\left((\Delta_0^2+\Delta_1^2)^2+\frac{(t_0^2+t_1^2)^2}{4}+(\Delta_0^2-\Delta_1^2)(t_1^2-t_0^2)+4\Delta_0\Delta_1t_0t_1\cos(\varphi)\right)^{\frac{1}{2}}\right)^{\frac{1}{2}} \\
                E_{even,7}^{(1)} &= -\left(\Delta_0^2+\Delta_1^2+\frac{t_0^2+t_1^2}{2}-\left((\Delta_0^2+\Delta_1^2)^2+\frac{(t_0^2+t_1^2)^2}{4}+(\Delta_0^2-\Delta_1^2)(t_1^2-t_0^2)+4\Delta_0\Delta_1t_0t_1\cos(\varphi)\right)^{\frac{1}{2}}\right)^{\frac{1}{2}}
\end{align}

Furthermore, we find:

\begin{align}
    H_{even}^{(2)}=\begin{pmatrix}0 & t_{0} & 0\\t_{0} & 0 & t_{1}\\0 & t_{1} & 0\end{pmatrix},
\end{align}
spanned by $\{|0\mathord{\da}\mathord{\da}\rangle,|\mathord{\da}0\mathord{\da}\rangle,|\mathord{\da}\mathord{\da}0\rangle\}$, and finally:

\begin{align}
H_{even}^{(3)}=\begin{pmatrix}0 & t_{0} & 0\\t_{0} & 0 & t_{1}\\0 & t_{1} & 0\end{pmatrix}
\end{align}

in the basis $\{|0\mathord{\ua}\mathord{\ua}\rangle,|\mathord{\ua}0\mathord{\ua}\rangle,|\mathord{\ua}\mathord{\ua}0\rangle\}$.
Note that $H_{even}^{(2)}$ and $H_{even}^{(3)}$ are the same and related by time-reversal symmetry.
Consequently, the eigenvalues are the same for each block reading

\begin{align}
    E^{(2,3)}_{even,1} &= 0, \\
    E^{(2,3)}_{even,2} &= \sqrt{t_0^2 + t_1^2}, \\
    E^{(2,3)}_{even,3} &= -\sqrt{t_0^2+t_1^2}. \\
\end{align}

\subsubsection{Odd parity sector}

In the odd parity sector, the Hamiltonian can similarly be split into three subblocks which can be solved separately.
Due to time reversal symmetry, the whole spectrum of the odd parity sector will be two-fold degenerate.
The first subblock reads:

\begin{align} \label{eq:odd_block_1}
    H_{odd}^{(1)}=\begin{pmatrix}
        0 & \Delta_{0} & 0 & 0 & 0 & 0\\\Delta_{0} & 0 & t_{1} & - \Delta_{0} & 0 & 0\\0 & t_{1} & 0 & 0 & t_{0} & 0\\0 & - \Delta_{0} & 0 & 0 & \Delta_{1} e^{i \varphi_{1}} & 0\\0 & 0 & t_{0} & \Delta_{1} e^{- i \varphi_{1}} & 0 & - \Delta_{1} e^{- i \varphi_{1}}\\0 & 0 & 0 & 0 & - \Delta_{1} e^{i \varphi_{1}} & 0
    \end{pmatrix},
\end{align}

in the basis $\{|\mathord{\ua}\mathord{\da}\mathord{\da}\rangle, |00\mathord{\da}\rangle, |0\mathord{\da}0\rangle, |\mathord{\da}00\rangle, |\mathord{\da}\mathord{\ua}\mathord{\da}\rangle, |\mathord{\da}\mathord{\da}\mathord{\ua}\rangle\}$ yields eigenenergies:

\begin{align}
    E_{odd,1}^{(1)} &= 0 \\
    E_{odd,2}^{(1)} &= 0 \\
    E_{odd,3}^{(1)} &= \left(\Delta_0^2+\Delta_1^2+\frac{t_0^2+t_1^2}{2}+\left((\Delta_0^2+\Delta_1^2)^2+\frac{(t_0^2+t_1^2)^2}{4}`+(\Delta_0^2-\Delta_1^2)(t_1^2-t_0^2)-2\Delta_0\Delta_1t_0t_1\cos(\varphi)-3\Delta_0\Delta_1\right)^{\frac{1}{2}}\right)^{\frac{1}{2}} \\
    E_{odd,4}^{(1)} &= -\left(\Delta_0^2+\Delta_1^2+\frac{t_0^2+t_1^2}{2}+\left((\Delta_0^2+\Delta_1^2)^2+\frac{(t_0^2+t_1^2)^2}{4}`+(\Delta_0^2-\Delta_1^2)(t_1^2-t_0^2)-2\Delta_0\Delta_1t_0t_1\cos(\varphi)-3\Delta_0\Delta_1\right)^{\frac{1}{2}}\right)^{\frac{1}{2}} \\ 
    E_{odd,5}^{(1)} &= \left(\Delta_0^2+\Delta_1^2+\frac{t_0^2+t_1^2}{2}-\left((\Delta_0^2+\Delta_1^2)^2+\frac{(t_0^2+t_1^2)^2}{4}`+(\Delta_0^2-\Delta_1^2)(t_1^2-t_0^2)-2\Delta_0\Delta_1t_0t_1\cos(\varphi)-3\Delta_0\Delta_1\right)^{\frac{1}{2}}\right)^{\frac{1}{2}} \\
    E_{odd,6}^{(1)} &= -\left(\Delta_0^2+\Delta_1^2+\frac{t_0^2+t_1^2}{2}-\left((\Delta_0^2+\Delta_1^2)^2+\frac{(t_0^2+t_1^2)^2}{4}`+(\Delta_0^2-\Delta_1^2)(t_1^2-t_0^2)-2\Delta_0\Delta_1t_0t_1\cos(\varphi)-3\Delta_0\Delta_1\right)^{\frac{1}{2}}\right)^{\frac{1}{2}}.
\end{align}

The time reversal symmetric block, written in the basis $\{|\mathord{\da}\mathord{\ua}\mathord{\ua}\rangle, |00\mathord{\ua}\rangle, |0\mathord{\ua}0\rangle, |\mathord{\ua}00\rangle, |\mathord{\ua}\mathord{\da}\mathord{\ua}\rangle, |\mathord{\ua}\mathord{\ua}\mathord{\da}\rangle$ has the same form as eq. \eqref{eq:odd_block_1}.
Consequently, the eigenvalues are the same as for the first block.
The third and final block written in the basis $\{|\mathord{\ua}\mathord{\ua}\mathord{\ua}\rangle, |\mathord{\da}\mathord{\da}\mathord{\da}\rangle\}$ is diagonal and features two eigenvalues:

\begin{align}
    E_{odd,1}^{(3)} = 0 \\
    E_{odd,2}^{(3)} = 0
\end{align}

\subsubsection{Analytic degeneracy condition}

Finally, we are interested in the condition yielding degeneracy of the groundstates.
To achieve this, we solve :
\begin{align}
    E_{odd,6}^{(1)} = E_{even,5}^{(1)}
\end{align}

for a relation between $t_i, \Delta_i$. 
Reducing the equations, we ultimately arrive at: 

\begin{align} \label{eq:degeneracy_asymmetric}
    \Delta_0\Delta_1 = -2t_0t_1\cos(\varphi),
\end{align}

which corresponds to eq. \eqref{eq:magic_line} given in the main text for $\Delta_0=\Delta_1$ and $t_0=t_1$.
Plugging eq. \eqref{eq:degeneracy_asymmetric} back into the previously calculated eigenvalues, we find that the spectrum becomes at least triply degenerate for each manifold.

\subsection{Insensitivity to spin-orbit interaction} \label{app:so_insensitivity}

In the theoretical analysis we make use of the fact that we can neglect spin-orbit interaction in our considerations.
We will demonstrate this assertion by explicitly constructing the gauge transformation that diagonalizes the hopping terms and show that the resulting Hamiltonians feature the same spectra in the new basis.
In the $U\rightarrow \infty$ limit the problem effectively becomes a single particle problem such that a treatment in the Bogoliubov-de Gennes basis becomes possible.
We write:

\begin{align}
H_{BdG} = \begin{pmatrix}\vec{c_1}^\dagger \\ \vec{c_2}^\dagger \\ \vec{c_3}^\dagger \\ \vec{c_4}^\dagger \\ \vec{c_5}^\dagger\end{pmatrix} \underbrace{\begin{pmatrix} H_1 & T_{1} & 0 & 0 & 0 \\ 
T_1^\dagger & H_{\mathrm{ABS},1} & T_2 & 0 & 0 \\ 0 & T_2^\dagger & H_2 & T_3 & 0 \\ 0 & 0 & T_3^\dagger & H_{\mathrm{ABS},2} & T_4 \\ 0 & 0 & 0 & T_4^\dagger & H_3\end{pmatrix}}_{h_{BdG}}  \begin{pmatrix}\vec{c_1}\\ \vec{c_2} \\ \vec{c_3} \\ \vec{c_4} \\ \vec{c_5}\end{pmatrix},
\end{align}

where $\vec{c_i}=(c_{i\uparrow}, c_{i,\downarrow}, c_{i\downarrow}^\dagger, c_{i\uparrow}^\dagger)^T$,
\begin{align}
H_i &= \mathrm{diag}[\mu_i, \mu_i, -\mu_i, -\mu_i] \\ 
H_{\mathrm{ABS},i} &= \begin{pmatrix} \mu_i & 0 & \Delta e^{i\varphi_i} & 0 \\ 0 & \mu_i & 0 & -\Delta e^{i\varphi}\\ \Delta e^{-i\varphi} & 0 & -\mu_i & 0 \\ 0 & -\Delta e^{-i\varphi} & 0 & -\mu_i\end{pmatrix},
\end{align}

and the hopping terms are given by:

\begin{align}
T_i = \begin{pmatrix} t_i & -t_{so,i} & 0 & 0 \\ t_{so,i} & t_i & 0 & 0 \\ 0 & 0 & -t_i & -t_{so,i} \\ 0 & 0 & t_{so,i} & -t_i\end{pmatrix} = |t_i|\underbrace{\begin{pmatrix} \cos(\theta_i) & -\sin(\theta_i) & 0 & 0 \\ \sin(\theta_i) & \cos(\theta_i) & 0 & 0 \\ 0 & 0 & -\cos(\theta) & -\sin(\theta_i) \\ 0 & 0 & \sin(\theta_i) & -\cos(\theta_i)\end{pmatrix}}_{U_i},
\end{align}

As we restrict the hopping to nearest neighbour only, we can perform a sequence of local gauge transformations on the fermionic operators to diagonalize the hopping terms.
Considering the transformation:

\begin{align}
    h_{BdG} = \begin{pmatrix}1&0 &0 & 0 & 0  \\0&U_1^\dagger &0 & 0 & 0  \\0&0 &1 & 0 & 0  \\0&0 &0 & 1 & 0  \\0&0 &0 & 0 & 1  \end{pmatrix} 
    \begin{pmatrix} H_1 & |t_1| & 0 & 0 & 0 \\ 
|t_1| & U_1H_{\mathrm{ABS}}U_1^\dagger & U_1T_2 & 0 & 0 \\ 0 & T_2^\dagger U_1^\dagger & H_2 & T_3 & 0 \\ 0 & 0 & T_3^\dagger & H_{\mathrm{ABS},2} & T_4 \\ 0 & 0 & 0 & T_4^\dagger & H_3\end{pmatrix}
 \begin{pmatrix}1&0 &0 & 0 & 0  \\0&U_1 &0 & 0 & 0  \\0&0 &1 & 0 & 0  \\0&0 &0 & 1 & 0  \\0&0 &0 & 0 & 1  \end{pmatrix} 
\end{align}

which diagonalizes $T_1$.
In the next step, the hopping between $H_{\mathrm{ABS},1}$ and $H_2$ can be diagonalized by pulling out unitary $\mathrm{diag}[1,1,U_1U_2,1,1]$.
Sequentially eliminating all spin-orbit unitaries leaves:

\begin{align}
    h_{BdG} =  U_{SO}^\dagger
    \begin{pmatrix} H_1 & |t_1| & 0 & 0 & 0 \\ 
    |t_1| & U_1H_{\mathrm{ABS},1}U_1^\dagger & |t_2| & 0 & 0 \\ 0 & |t_2| & (U_1U_2)H_2(U_1U_2)^\dagger & |t_3| & 0 \\ 0 & 0 & |t_3| & (U_1U_2U_3)H_{\mathrm{ABS},2}(U_1U_2U_3)^\dagger & |t_4| \\ 0 & 0 & 0 & |t_4| & (U_1U_2U_3U_4)H_3(U_1U_2U_3U_4)^\dagger\end{pmatrix}
    U_{SO},
\end{align}

with:

\begin{align}
    U_{\mathrm{SO}}=\begin{pmatrix}1&0 &0 & 0 & 0  \\0&U_1 &0 & 0 & 0  \\0&0 &U_1U_2 & 0 & 0  \\0&0 &0 & U_1U_2U_3 & 0  \\0&0 &0 & 0 & U_1U_2U_3U_4  \end{pmatrix} .
\end{align}

Since $H_i = \mathrm{diag}[\mu_i, \mu_i, -\mu_i, -\mu_i]$, it can be seen that the normal dots remain invariant by this transformation and are hence insensitive to spin-orbit.
Similarly, one finds that the Bogoliubov spectrum remains invariant by the unitaries:

\begin{align}
    U_iH_{ABS,i}U_i^\dagger & = \begin{pmatrix} t_i & -t_{so,i} & 0 & 0 \\ t_{so,i} & t_i & 0 & 0 \\ 0 & 0 & -t_i & -t_{so,i} \\ 0 & 0 & t_{so,i} & -t_i\end{pmatrix} \begin{pmatrix}
    \mu_i & 0 & \Delta e^{i\varphi} & 0 \\ 
    0 & \mu_i & 0 & -\Delta e^{i\varphi} \\ 
    \Delta e^{-i\varphi} & 0 & -\mu_i & 0 \\
    0 & -\Delta e^{-i\varphi} & 0 & -\mu_i
    \end{pmatrix}
    \begin{pmatrix} t_i & t_{so,i} & 0 & 0 \\ -t_{so,i} & t_i & 0 & 0 \\ 0 & 0 & -t_i & t_{so,i} \\ 0 & 0 & -t_{so,i} & -t_i\end{pmatrix}  \nonumber \\
    &=  \begin{pmatrix}
    \mu_i & 0 & -\Delta e^{i\varphi} & 0 \\ 
    0 & \mu_i & 0 & \Delta e^{i\varphi} \\ 
    -\Delta e^{-i\varphi} & 0 & -\mu_i & 0 \\
    0 & \Delta e^{-i\varphi} & 0 & -\mu_i
    \end{pmatrix},
\end{align}

where we have used that $t^2+t_{so}^2=1$.
While the pairing term changes its sign, the spectrum remains invariant implying that the effective model remains the same.
Lastly, because $T_i$ is a block-diagonal consisting of rotation matrices we can deduce from the group properties of rotation matrices that this invariance remains also true for products of $U_i$ transforming a $H_{\mathrm{ABS},j}$.

\newpage
\section*{Supplementary Figures S1 to S9}

\begin{figure}[h!]
\centering
    \includegraphics[width = \textwidth]{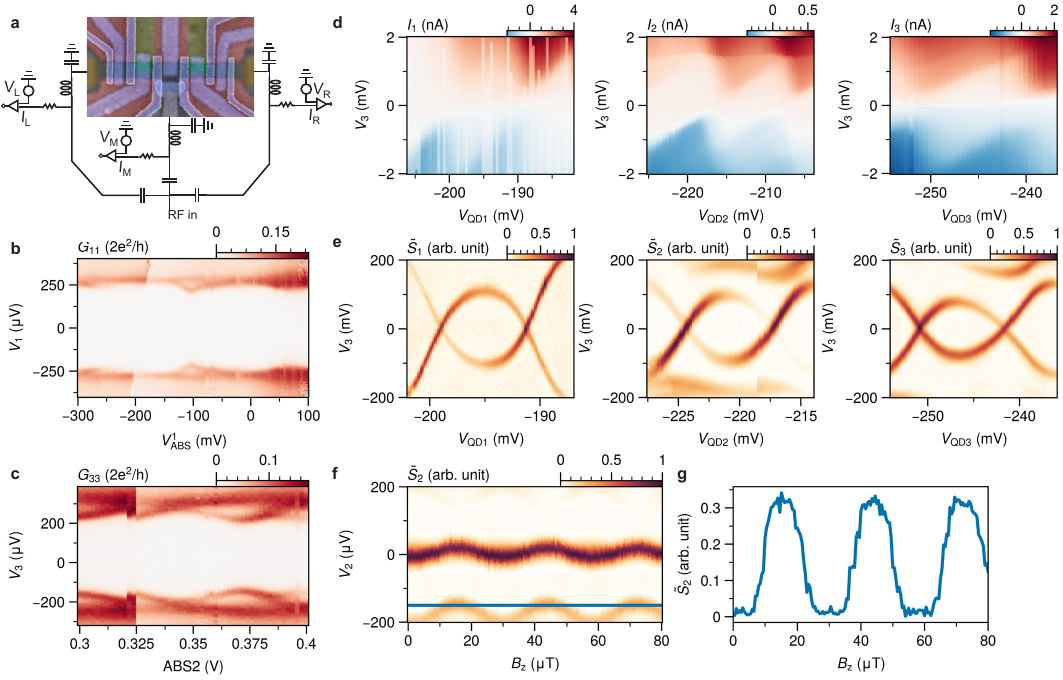}
    \caption{\textbf{Characterisation device A} 
    \fl{a} SEM of a copy of device A, used to obtain the results presented in the main text. 
    The RF and DC circuit components are included. 
    As first characterisation, the induced superconductivity in the hybrid sections is investigated by activating only the tunneling gates next to each aluminium strip.
    We then obtain finite-bias spectroscopy of \fl{b} the left hybrid segment while sweeping \ABS{1} and \fl{c} the right hybrid segment while sweeping \ABS{2}. 
    In both sides, an induced superconducting gap around \SI{250}{\micro\volt} is observed.
    \fl{d} Next, all tunneling gates are activated to define three QDs. 
    To allow for strong interactions between them, the tunneling barriers between each QD and their neighbouring hybrid segments are kept relatively open.
    In this regime, no fully Coulomb-blockaded diamonds are observed in the currents measured in (d), obtained with the QD settings used for measurements obtained in the main text .
    Nevertheless, one can roughly estimate the charging energies in each QD to be on the order of \SI{1.5}{\milli\volt}.
    \fl{e} Lower-bias spectroscopy in the same range as the measurements in (d), showing sub-gap states with an eye-shaped dispersion as a function of the QD plunger gates, typical for proximitized QDs.
    \fl{f} The proximity to two SC leads additionally results in the middle QD's spectrum becoming sensitive to the flux threaded through the SC loop. 
    This effect is corrected in the measurements where the phase is varied, by retuning the QD to charge degeneracy.
    \fl{g} Additionally, the oscillations allows for estimating the SC period to be \SI{28}{\micro\tesla}, which we use to map \var{B}{z} to \var{\phi}.
    }
    \label{fig:S1_charact_devA}
\end{figure}

\begin{figure}[h!]
\centering
    \includegraphics[width = \textwidth]{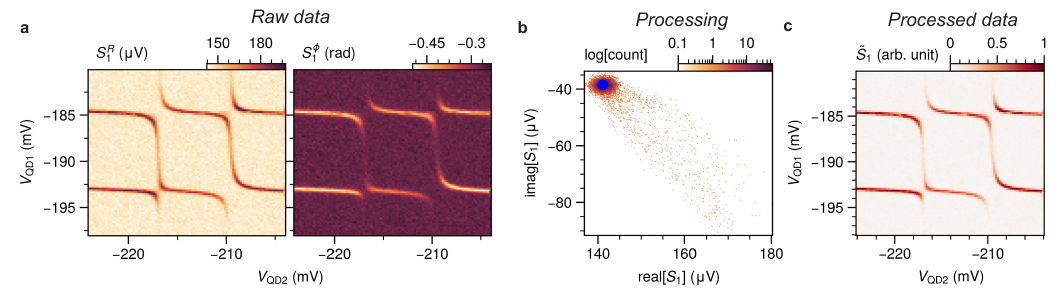}
    \caption{\textbf{Example of RF-signal processing.} In lead-reflectometry measurements, both the amplitude ($S_{\mathrm{i}}^{\mathrm{R}}$) and phase $S^{\phi}_{\mathrm{i}}$ of the reflected signals is recorded (e.g. $S_{\mathrm{1}} = S_{\mathrm{1}}^{\mathrm{R}} e^{\mathrm{i} S^{\phi}_{\mathrm{1}}}$). 
    We convert this to a single value to display in the main text, visualised here.
    \fl{a} Raw data corresponding to the measurement shown in \ref{fig:Fig2}a. 
    Both the amplitude response (left) and the phase response (right) are recorded.
    \fl{b} Visualisation of the data processing. 
    First, the complex values $S_1$ are collected in a 2-d histogram on the complex plane. 
    The value corresponding to Coulomb blockade will show up in this plane as the point with the highest count, indicated here with the blue mark.
    Consequently we convert each point to the processed data-points $\tilde{S}_1$, defined as the distance of each point to Coulomb blockade.
    An example for a single data-point is highlighted with the arrow.
    \fl{c} Processed data of (a), identical to the measurement shown in \cref{fig:Fig2}a.
    }
    \label{fig:S2_RF_processing}
\end{figure}

\begin{figure}[h!]
\centering
    \includegraphics[width = \textwidth]{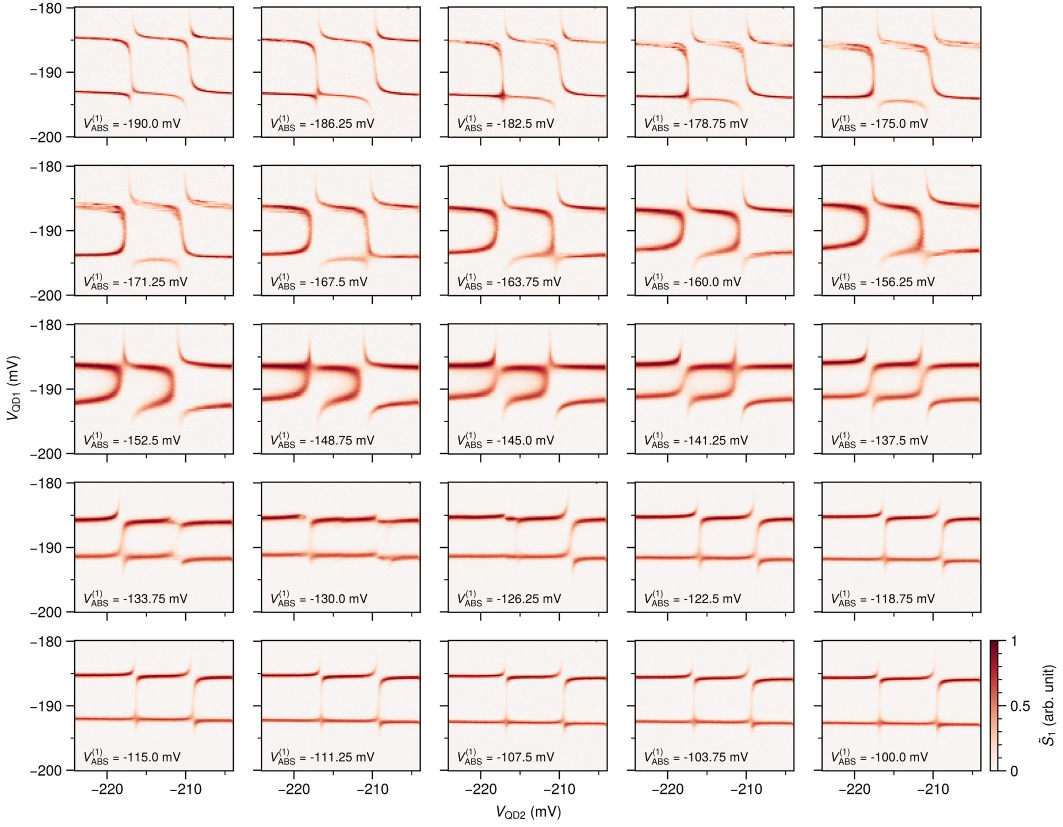}
    \caption{\textbf{CSDs in extended range for device A.} Main text \cref{fig:Fig2} highlights two examples of CSDs obtained for the three-site device, sweeping QD1 and QD2 when QD3 is kept off resonance. 
    Here an extended range is shown, for CSDs obtained for 25 values of \ABS{1} in the range \SI{-190}{\milli\volt} to \SI{-100}{\milli\volt}.
    In particular, these measurements highlight how each quadrant transitions from an ECT dominated to a CAR dominated avoided crossing, indicating that a sweet spot can be reached for each charge configuration in this range.}
    \label{fig:S3_extended_CSDs_devA}
\end{figure}

\begin{figure}[h!]
\centering
    \includegraphics[width = \textwidth]{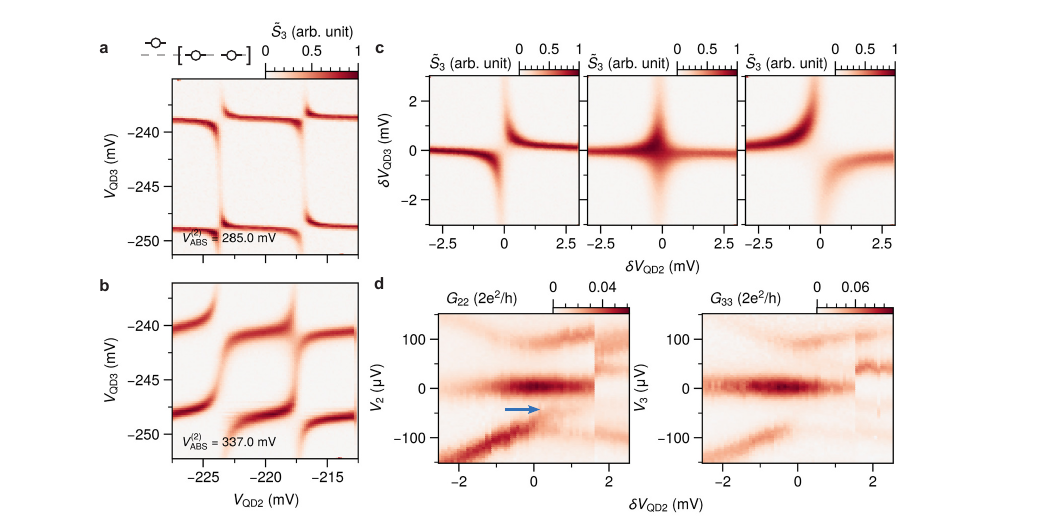}
    \caption{\textbf{Characterisation of right PMM pair in device A.} 
    In the main text, the left and middle QDs are used to demonstrate a sweet spot in the effective two-site chain. 
    To perform the measurements in Figs.~3-5, the pair formed by the middle and right QD must similarly be at a two-site sweet spot.
    To find this, the same tuning procedure is used as for the left pair in \cref{fig:Fig2}.
    \fl{a} CSDs obtained with the left QD tuned off resonance, at \ABS{2}~=~\SI{285}{\milli\volt}, showing an ECT dominated avoided crossing in each quadrant.
    \fl{b} CSDs obtained at \ABS{2}~=~\SI{337}{\milli\volt}, now showing CAR dominated avoided crossings. 
    The transition shows that a sweet spot can also be found for each quadrant in this right QD.
    \fl{c} Close-ups of the bottom left quadrant, fine-tuning the ABS
    \fl{d} Finite-bias spectroscopy at the sweet-spot in (c), while sweeping \var{V}{QD2} along the highlighted path.
    The triplet feature distinguishing these measurements from the high-field case is again (faintly) visibly, indicated by the arrow.
    }
    \label{fig:S4_right_PMM}
\end{figure}

\begin{figure}[t!]
\centering
    \includegraphics[width = \textwidth]{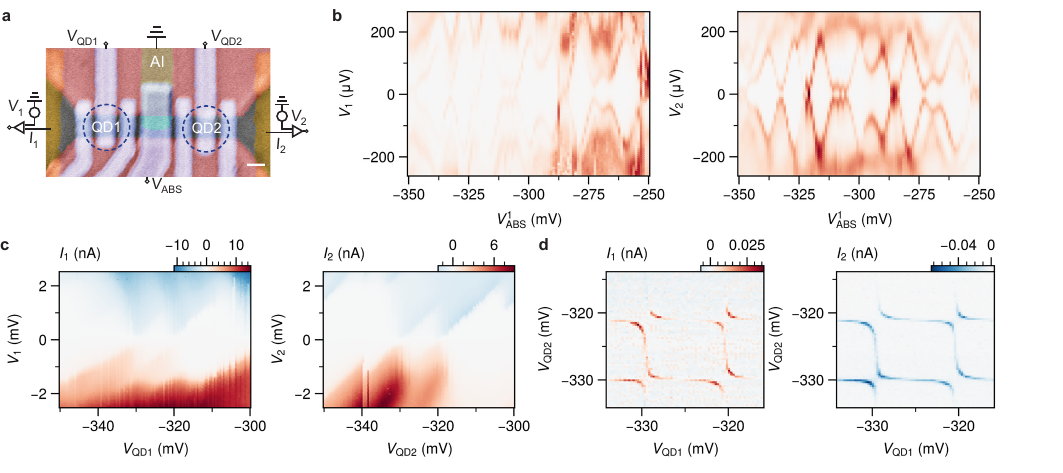}
    \caption{\textbf{Device B: Basic characterisation.}  
    \fl{a} False colour scanning electron micrograph of device B.
    \fl{b} Characterisation of the the hybrid section, through finite-bias spectroscopy measured from the left and right sides.
    The spectrum of Andreev bound states is more crowded and shows a larger charging energy, compared to the sections in device A (\ref{fig:S1_charact_devA}).
    Nonetheless, suitable regions in parameter space of \ABS{1} could be located. 
    \fl{c} Measurements of Coulomb diamonds in QD1 and QD2, that are used for the supplementary data in Figs.~\ref{fig:S6_extended_CSDs_devB},\ref{fig:S7_reproduce_spectra_devB}.
    \fl{d} Exemplar charge stability diagram for a pair of resonances in QD1 and QD2, with avoided crossings here indicating strong ECT interactions between the QDs.
    }
    \label{fig:S5_devBcharacterisation}
\end{figure}

\begin{figure}[t!]
\centering
    \includegraphics[width = \textwidth]{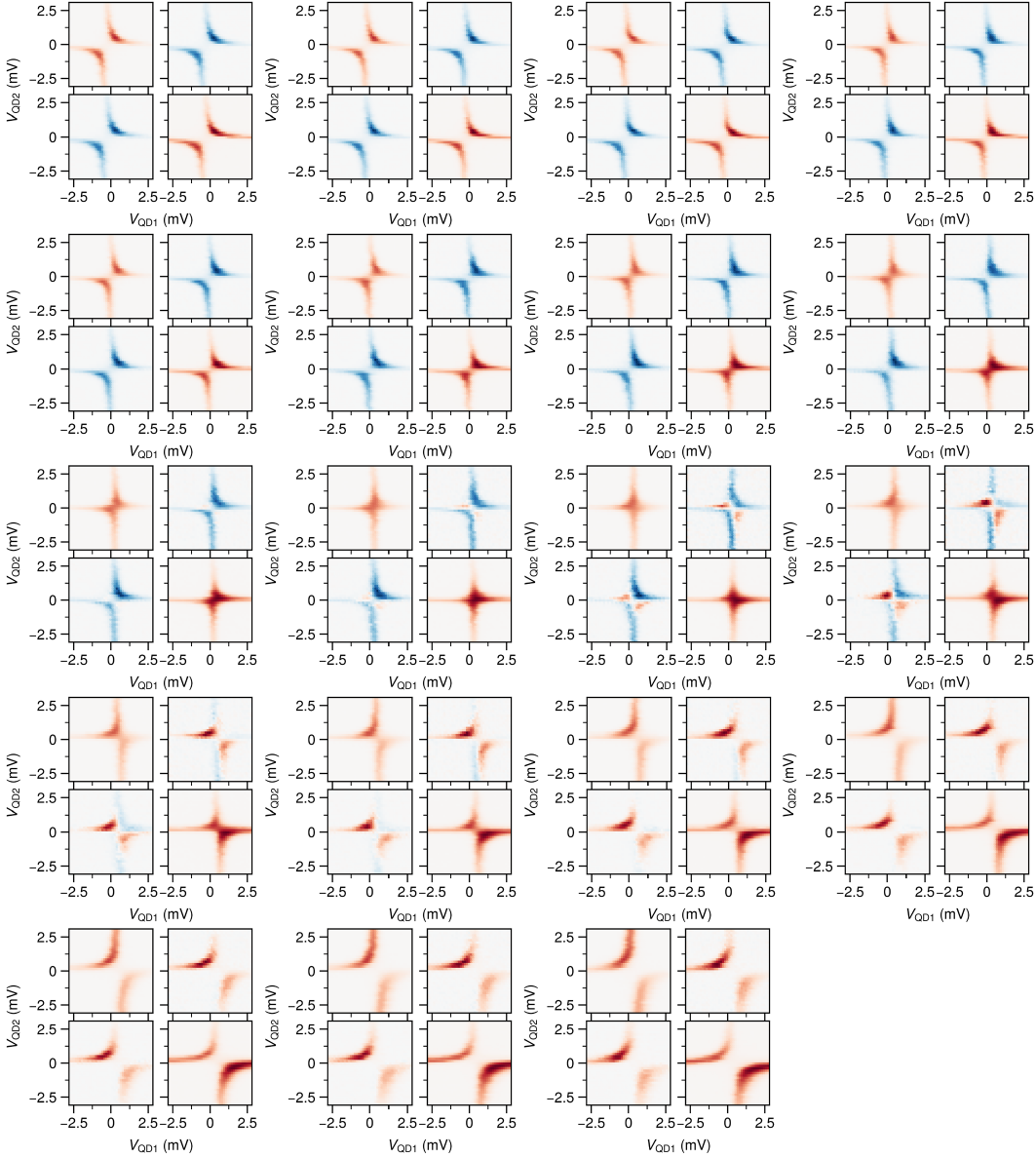}
    \caption{\textbf{Device B: charge stability diagrams in a wider \ABS{1} range.} A key finding in \cite{tenHaaf2024}, was that the CSDs for a two-site chain at a sweet spot at finite field, could not be distinguished from the CSDs at zero magnetic field, contrasting predictions~\cite{Tsintzis20222} that the non-local conductance could be used to determine the quality of Majorana's. 
    Here we present local and non-local zero-bias conductance CSDs in device B, for a more extensive range of \ABS{1} around a two-site sweet spot. 
    The lower left charge configuration from \ref{fig:S5_devBcharacterisation}d is shown.
    In particular, as the system transition from an ECT dominated to a CAR dominated regime, a switch is observed between measuring negative non-local conductance to measuring positive non-local conductance.
    At the sweet spot, crossing forms around \var{\mu}{i}=0 where the non-local conductance is zero, owed to the chargeless nature of the excitations at the sweet spot.
    Notably, the entire trend shown here is identical to measurements obtained at finite magnetic fields~\cite{Dvir2023}.
    }
    \label{fig:S6_extended_CSDs_devB}
\end{figure}

\begin{figure}[t!]
\centering
    \includegraphics[width = \textwidth]{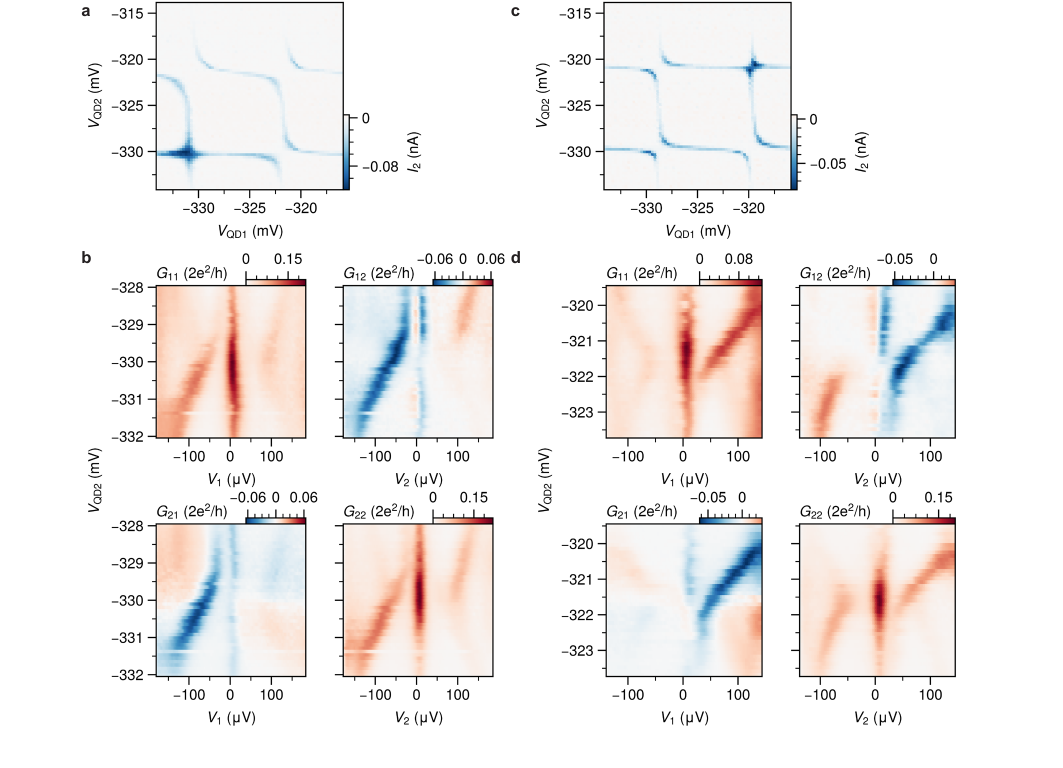}
    \caption{\textbf{Device B: conductance spectra and role of charge configuration.} In main text \cref{fig:Fig2}, the conductance spectra of the two-site sweet spot as a function of QD plunger gate is presented. 
    In particular, such a measurement reveals the only signature that distinguishes the zero-field and finite-field systems: a particle-hole symmetry breaking feature related to the presence of triplet states in the even subspace.
    Interestingly, the feature can appear either on the hole-side or electron-side of the applied voltage bias, depending on the charge configurations of the QDs.
    We reproduce these measurements here for two charge configurations in Device B. 
    \fl{a} CSD taken at a voltage \ABS{1} that corresponds to a sweet spot in the lower left quadrant.
    \fl{b} Full conductance matrix measured as a function of detuning \var{V}{QD2}, along the dashed line in (a). 
    The triplet feature connecting the excited states and the zero energy states appears at negative bias.
    \fl{c} CSD taken at a voltage \ABS{1} corresponding to a sweet spot in the top right quadrant.
    \fl{d} Full conductance matrix measured as a function of detuning \var{V}{QD2}, along the dashed line in (c). 
    The triplet feature again appears, but now at positive bias, as the excitation has now become `electron-like' due to the new charge configuration of the QDs.
    }
    \label{fig:S7_reproduce_spectra_devB}
\end{figure}

\begin{figure}[t!]
\centering
    \includegraphics[width = \textwidth]{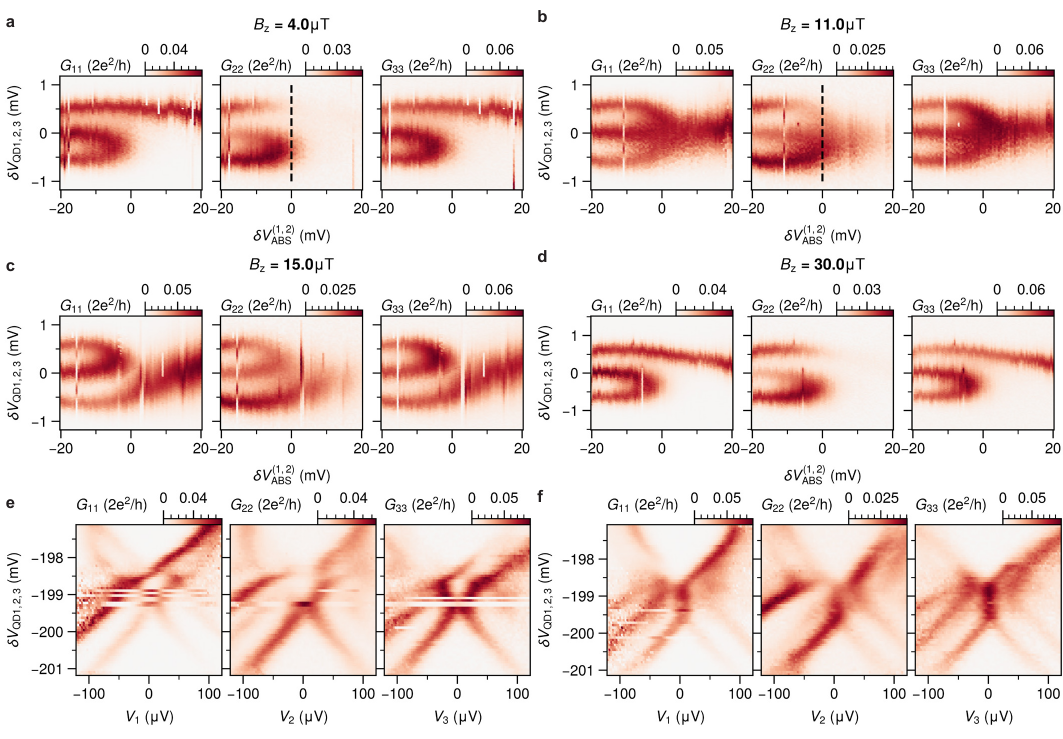}
    \caption{\textbf{Raw datasets for figure 4.} Main text \cref{fig:Fig4} displays zero-bias conductance measurements obtained in a large parameter space of sweeping simultaneously all QD plungers versus sweeping both \ABS{1} and \ABS{2} around a sweet spot value (denoted $\delta$\ABS{1}=0). 
    In the main text, only \var{G}{11} is shown, while \var{G}{22} and \var{G}{33} were measured simultaneously. 
    We show here the full dataset, for \fl{a} \var{B}{z}~=~\SI{4}{\micro\tesla}, \fl{b} \var{B}{z}~=~\SI{11}{\micro\tesla}, \fl{c} \var{B}{z}~=~\SI{15}{\micro\tesla} and  \fl{d} \var{B}{z}~=~\SI{30}{\micro\tesla}.
    In addition, finite bias spectroscopy was obtained along the dashed lines in (a) and (b), when simultaneously sweeping all QD plunger gates, shown in (e) and (f) respectively.
    At the \var{B}{z} value corresponding to the `special angle' in (b), the zero-bias conductance appears to split non-linearly, reminiscent of the behaviour expected for this system at finite field. 
    }
    \label{fig:S8_full_data_figure_4}
\end{figure}

\begin{figure}[t!]
\centering
    \includegraphics[width = \textwidth]{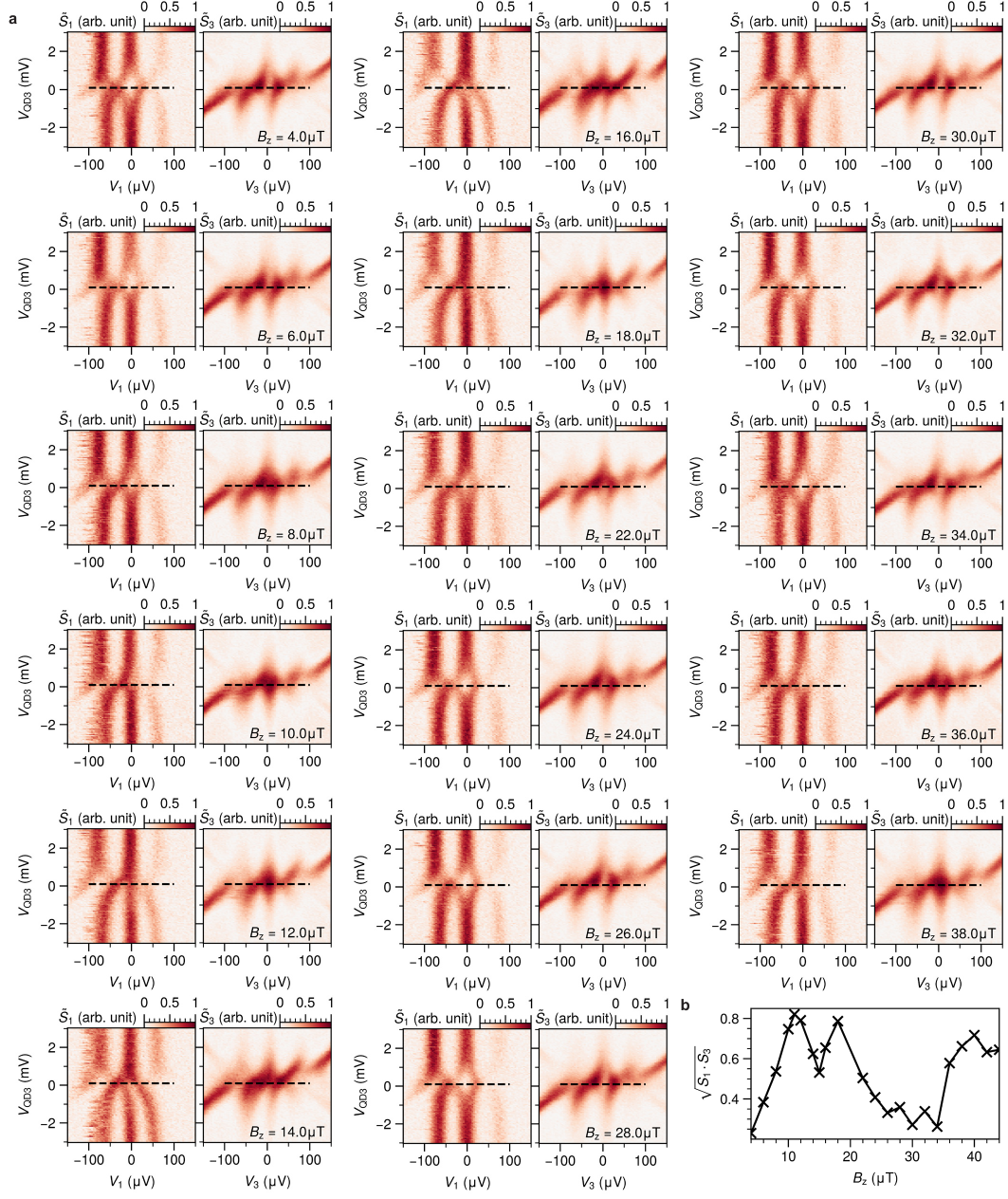}
    \caption{\textbf{Additional datasets for figure 5}  \fl{a} In main text \cref{fig:Fig5}f, a finite-bias spectroscopy versus detuning \var{V}{QD3} is shown for a value of \var{B}{z} to match the theoretical angle where a seemingly stable zero-bias peak appears. 
    Here, we show the evolution of this measurement as a function of \var{B}{z},observing the smooth transition from \cref{fig:Fig3}a to \cref{fig:Fig5}f.
    \fl{b} To visualize the trend in closing and re-opening of the splitting of the ZBP as \var{V}{QD3} is brought on resonance, we extract $\sqrt{\tilde{S}_1\cdot\tilde{S}_2}$ at \var{\delta V}{QD3}, \var{V}{1} and \var{V}{3} = 0.
    }
    \label{fig:S9_detuning_vs_phase}
\end{figure}


\end{document}